\documentclass[preprint2]{aastex}

\newcommand{\EQ}{\begin{equation}}
\newcommand{\EN}{\end{equation}}
\newcommand{\EQA}{\begin{eqnarray}}
\newcommand{\ENA}{\end{eqnarray}}
\newcommand{\BB}{{\bf B}}
\newcommand{\nab}{{\nabla}}
\newcommand{\uu}{{\bf u}}
\newcommand{\rrr}{{\bf r}}

\newcommand{\rr}{{\bf {\hat r}}}

\newcommand{\OO}{\mbox{\boldmath $\Omega$} {}}

\shorttitle{Sun's preferred longitudes and dynamo}

\shortauthors{Bigazzi \& Ruzmaikin}

\begin{document}

\title{The Sun's Preferred Longitudes and the Coupling of Magnetic Dynamo Modes.}

   \author{Alberto Bigazzi and Alexander Ruzmaikin}
     \affil{Jet Propulsion Laboratory, California Institute of Technology,
     CA USA}

   \begin{abstract}
Observations show that solar activity is distributed non-axisymmetrically, concentrating at ``preferred longitudes".  This indicates the important role of non-axisymmetric magnetic fields in the origin of solar activity.  We investigate the generation of the non-axisymmetric fields and their coupling with
axisymmetric solar magnetic field.  Our kinematic generation (dynamo) model operating in a sphere
includes solar differential rotation, which approximates the differential rotation obtained by inversion of helioseismic data, modelled distributions of the turbulent resistivity, non-axisymmetric mean helicity,
and meridional circulation in the convection zone.  We find that 
(1) the non-axisymmetric modes are localised near the base of the convection zone,  where the formation of active regions starts,  and at latitudes around $30^{\circ}$; 
(2) the coupling of non-axisymmetric and axisymmetric modes causes the non-axisymmetric mode to follow the solar cycle;  the phase relations between the modes are found. 
(3) The rate of rotation of the first non-axisymmetric mode is close to that
determined in the interplanetary space. 
     \end{abstract}

\section{Introduction}
Solar magnetic activity tends to cluster at ``preferred longitudes''
\cite{vitin69,bogart82, bai87,antonucci90,benevolen99,DeTomaEtal2000}. 
Spacecraft data collected over
three solar cycles indicate a persistent dependence of the solar wind
speed and the radial component of the interplanetary magnetic field
on solar longitude. 
Patterns are found to be 
rotating with a fixed period \cite{neugebauer00}. 
This indicates the involvement of non-axisymmetric 
(i.e. longitude dependent) large-scale magnetic fields in the 
formation and evolution of  solar activity as seen in  
solar magnetograms \cite{ruzmaikin01b,henney03}. 

The direct expansion of observed solar magnetograms into spherical
harmonics shows the presence of axisymmetric (longitude-independent)
and non-axisymmetric modes 
\cite{altschuler74,DeTomaEtal2000}. The axisymmetric
mode dominates at solar minimum and the first non-axisymmetric mode is
prevalent at solar maximum  \cite{ruzmaikin01b}. Earlier studies
\cite{stix74,KR80,ivanova85} have shown that non-axisymmetric modes could be generated by the mean-field dynamo under certain constraints imposed on 
the distributions of the differential rotation
and the mean kinetic helicity ($\alpha$-effect). 
These earlier studies did not include the correct distribution, 
in radius and latitude, of the solar differential rotation curve, 
unknown at the time. A more recent model of Moss (1999), which includes the 
updated rotation curve, also displays the generation of a 
non-axisymmetric component of the field. 

The non-axisymmetric magnetic fields are generated by axisymmetric 
sources but, in kinematic mean-field dynamos, 
evolve  independently of the axisymmetric field.
In particular, the non-axisymmetric modes
do not have the same period as the axisymmetric mode. 
However, the observation that in the Sun the non-axisymmetric fields
do have the same solar cycle as the axisymmetric fields,  
suggests a coupling between the modes that  
locks the oscillation  periods of the two together. 
Here we prove that this is in fact possible and it can be achieved  by
relaxing the assumption of axisymmetry for the $\alpha$-effect. 

The coupling of different modes by non-axisymmetric motions
was first studied for the Earth's magnetic fields by Bullard \& Gellman (1954). 
% CHANGED 
Non-axisymmetry is a key ingredient in models of tachocline solar flux tube
dynamics that studied the instabilities leading to rise of the flux
tubes \cite{Choudhuri89,FanEtal93,FerrizSchue94, caligari95}. In these models the
formation of magnetic loops emerging at the solar surface as active
regions is considered to be a consequence of these non-axisymmetric
(kink-type) instabilities. Although the dynamo itself is not treated in these studies, 
longitudinal instabilities are connected to the emergence of active regions, see also Ruzmaikin (2001).
A calculation of the $\alpha$-effect from these instabilities,  have been 
performed by \cite{FerrizShmSchue94}.
  Such an $\alpha$-effect, however, 
 has not yet  been included in mean-field dynamo models. 
Kinetic helicity in hydrodynamic simulation of shallow water instabilities
used in  models of the tachocline, also reveal a non-axisymmetric, $m=1$, 
pattern for the $\alpha$-effect  \cite{DikGil01}. 
That is not included, however, in the  dynamo model that the same  authors
have analysed. 
A non-axisymmetric $\alpha$-effect has been considered in mean-field dynamo
models of the galactic dynamo \cite{moss91}. 

% CHANGED 
Here for the first time we study the solar mean-field dynamo with 
a non-axisymmetric $\alpha$.  
We discuss the distribution in radius latitude and azimuth 
of the first non-axisymmetric mode
and relate it to preferred longitudes and clustering of magnetic activity. 
It is important to note that,  contrary to the axisymmetric fields, 
the non axisymmetric modes are rotating structures in the frame of reference
of the body.
The rate of rotation of these modes is 
determined by the solution of the dynamo equations. 
We determine the rate for the $m=1$ mode by looking at the evolution of 
the surface patterns of its radial magnetic field. Because the surface poloidal field expands into the solar wind, we  relate  this rate  to the signatures of preferred longitudes found in  the solar wind.

Our kinematic model includes an approximation to the helioseismically observed distribution of the differential rotation and modelled distributions of other sources of the mean field generation. 

\section{The Non-Axisymmetric Dynamo Model}
\label{s_model}
When the motions are given, the generation of the mean magnetic field is
described by the equations
\EQ
\label{eq_induction}
\partial_t\BB= \nab\times\left(
\uu\times\BB+ \alpha\BB-\eta \BB
\right),
\EN
\EQ
\label{eq_divb}
\nab\cdot\BB=0
\EN
\cite{moffatt78,Parker79,KR80}, 
where $\BB$ is the mean magnetic field (the first
statistical moment of the total field), $\uu$ is the mean velocity,
$\alpha$ is the mean-helicity, and $\eta$ is the electrical resistivity,
which includes the turbulent resistivity.  

Solar rotation is approximated by a simple analytical fit
\cite{charbonneau99,moss99}
to the distribution reconstructed by helioseismic methods
\cite{kosovichev96,howe00},
see Fig.~\ref{fig_rotOurs}. Details are given in the  Appendix.
The rate of the equatorial rotation is taken as
$\Omega_{eq}/2\pi=460.7~{\rm nHz}$ $=4.607\times10^{-9} \mbox{s}^{-1}$.
This distribution matches the core rotation  $\Omega_{\mbox{c}}$ smoothly
in a layer of thickness $0.04 R_{\odot}$
at $0.69 \mbox{R}_{\odot}$. 
The core rotation is about equal to the surface rotation  $\Omega_s $ at 
$30^{\circ}$ latitude. The surface shear layer is not included in this model.
However, the deep shear layer at $0.7 \mbox{R}_{\odot}$, i.e. the tachocline, is well approximated. 

The distribution of $\alpha$ cannot yet be determined from
observations. We therefore consider three typical   cases, see Appendix. 
The radial profiles are shown in Fig.~\ref{fig_alp}.
In the first case, M1, the distribution of $\alpha$ is independent of radius in the convection
zone and rapidly decreases below the convection zone. 
In the second case, M2, \(\alpha\) is concentrated near the solar surface and does not overlap
with the shear layer.  In the third case, M3,  \(\alpha\) is localised above the tachocline
\cite{charbonneau97,mason02} overlapping with the radial shear.
The latitudinal dependence of $\alpha$ in all three cases is assumed to be $\cos\theta$.
This simple angular dependence is partially justified by the fact that the
pseudo-scalar $\alpha$ can be formed from the scalar product of the
density gradient and rotation: \(\alpha \sim \nab\rho
\cdot \OO \sim \cos\theta \). More elaborate angular dependences are
produced by different models of $\alpha$, see for example 
%\cite{RueBran95,DikGil01} 
R\"udiger \& Brandenburg (1995), Dikpati \& Gilman
(2001), and can easily be incorporated into our model.

The resistivity $\eta$ is taken constant throughout the convection zone with 
a sharp decrease in the transition layer to the highly conductive radiative interior. We consider the case when the transition layer is coincident with the rotational shear layer. For computational needs,  we assume that the resistivity near the base of the convection zone is $50$ to $200$ times less than the turbulent resistivity in the bulk of the convection zone.  (A realistic number would be of the order of $10^6$). 
We use $R_{\odot}^2/\eta$ as the time unit.  We assign the value of the 
resistivity $\eta$ by  considering the marginal solution of model M1. 
We match the equatorial rotation rate in Fig.~\ref{fig_rotOurs} to the
observed one, obtaining  $\eta_0=2.74 \times 10^{11} {\rm cm}^2{\rm s}^{-1}$. 
Estimates made using classical mixing length theory,  
give  a value for the resistivity of order $ 10^{12}{\rm cm}^2{\rm s}^{-1}$. 

The ratio $T_{c}/T_{rot}$ of the cycle period
to the rotation rate of the body is
non-dimensional and does not therefore depend on the choice of $\eta$. 
For the Sun, the ratio of the cycle and the
equatorial rotation rate,  is approximately $150$. 
Higher values of this number imply a cycle period of the solutions 
longer than the solar one.  Shorter values, a shorter cycle period. 
We characterise the  cycle periods of our solution 
in terms of this non-dimensional ratio.
%, see  Table~\ref{tab_result}.

In the case of non-axisymmetric dynamo an analogous ratio of 
the rate of rotation of  the non-axisymmetric
mode to the body rotation rate could also be used 
to characterise the solutions. 
Such a number clearly does not exist in the case of the axisymmetric dynamo
where no rotating structures exist. 
%
%The values of this quantity for our three models are shown in Table~\ref{tab_result}. While for model M1 and M3 we have a very good agreement with the solar value, for model M2 this ratio is much smaller. 

\subsection{Mathematical Formulation and Numerical Procedure}
We have developed a new code that solves the kinematic dynamo equations
(\ref{eq_induction},\ref{eq_divb}) in a spherical domain. 
We define a regular grid in the meridional circular sector 
$r\in[r_{in},R_{\odot}],
\theta\in[0,\pi]$, where $\theta$ is the co-latitude.  
In the azimuthal direction $\phi$ we expand all functions in m-modes,
$e^{im\phi}$. 
Solving the equations in the real space has advantages over using
the spectral Legendre Transform. The m-modal transform can be numerically
performed in a very efficient way by means of the Fast Fourier Transform
algorithm. No such algorithm exists for the 
Legendre Transform, needed for the latitudinal direction $\theta$. This makes
spherical spectral codes numerically expensive to run. 
It is also easier to parallelise a code in a regular domain. 

We represent the divergence-free magnetic field by two scalar functions T and S, called the toroidal and poloidal potentials
\EQ
{\BB}=\underbrace{-\rrr\times{\nab} T}_{Toroidal}
+ \underbrace{\nab\times(-\rrr\times{\nab} S)}_{Poloidal}.
\EN
\cite{moffatt78,KR80}.
This representation ensures that the field is divergence free. 
The curl of a poloidal field is a toroidal field and vice versa.

The choice of  $T$ and $S$ is made unique by means of the gauge condition
\EQ
\label{eq_gauge}
\int_{\Sigma} T\; \sin\theta \; d\theta \;  d\phi \;   =0 \;,\;
\int_{\Sigma} S\; \sin\theta \; d\theta \;  d\phi \;   =0.
\EN
 When the field vanishes, $\BB =0$, then its toroidal and poloidal
components $\BB_p$ and $\BB_t$ vanish as well.  
The vector equation (\ref{eq_induction}) therefore reduces to two 
coupled equations for the toroidal and poloidal potentials.

When $\alpha = \alpha(r,\theta)$ and $\eta = \eta (r)$, the governing
Eqs (\ref{eq_induction}), in spherical coordinates $r, \theta, \phi$,
reduce to
\EQA
\label{eq_tp}
\partial_t T &=&
R_{\Omega} V_{\Omega} + R_{\alpha}V_{\alpha} + R_M V_M
\nonumber \\
&+& \eta \nabla^2T
+ \partial_r \eta \cdot \frac{1}{r}\partial_r (rT),
\nonumber \\
\partial_t S &=&R_{\Omega} U_{\Omega} + R_{\alpha} U_{\alpha} + R_M U_M
\nonumber \\
&+&\eta \nabla^2S
\ENA
with the gauge conditions (\ref{eq_gauge}).
The non-dimensional numbers $R_{\Omega}$, $R_{\alpha}$, $ R_M$ are defined as:
\EQ
R_{\Omega}= \frac{\Omega_0 R_{\odot}^2}{\eta_0}  , \;\;\;
R_{\alpha}=\frac{\alpha_0 R_{\odot}}{\eta_0} , \;\;\;
R_{M}=\frac{u_M R_{\odot}}{\eta_0}.
\EN
Subscripted quantities  indicate typical amplitudes.
The functions $U_{\Omega}, V_{\Omega}, U_{\alpha}, V_{\alpha}, U_M, V_M,$
 are related to the toroidal
and poloidal parts of the sources:
\EQA
((\OO\times  \rrr) \times\BB)_T &=& -\rrr\times\nab U_{\Omega},  \\
 \nab\times((\OO\times  \rrr) \times\BB)_P &=&
-\rrr\times\nab V_{\Omega}, \\
(\uu_{M}\times\BB)_T &=& -\rrr\times\nab U_{M},  \\
 \nab\times(\uu_{M}\times\BB)_P &=&
-\rrr\times\nab V_{M}, \\
(\alpha \BB)_T &=& -\rrr\times\nab U_{\alpha}, \\
\nab \times (\alpha \BB)_P &=& -\rrr\times\nab V_{\alpha}.
\ENA
We have split the velocity field $\uu$ into its rotational (toroidal) and
meridional components 
\EQ
\uu(r,\theta)=
\OO(r,\theta)\times  \rrr + \uu_M(r,\theta).
\EN

The  functions $U_{\epsilon}, V_{\epsilon}, U_{\alpha}, V_{\alpha}, U_M, V_M
$ can be calculated from the above  relations:
\EQA
\label{eq_rdotcurl}
-L^2 V_{\Omega} &=& 
\rrr\cdot \nab\times\nab\times((\OO\times  \rrr) \times\BB), \\
-L^2 U_{\Omega}&=&  \rrr\cdot \nab\times((\OO\times  \rrr)\times\BB), \\
-L^2 V_{M} &=& 
\rrr\cdot \nab\times\nab\times(\uu_{M}\times\BB), \\
-L^2 U_{M}&=&  \rrr\cdot \nab\times(\uu_{M}\times\BB), \\
-L^2 V_{\alpha}&=& \rrr\cdot \nab\times\nab\times(\alpha(r,\theta)\
\BB), \\
-L^2 U_{\alpha}&= & \rrr\cdot \nab\times(\alpha(r,\theta)\ \BB),
\label{eq_rdotcurl6}
\ENA
where
\EQ
L^2= {1\over\sin \theta} {\partial\over\partial\theta}
\left( \sin\theta {\partial\over\partial\theta} \right)
+ {1\over\sin^2\theta} {\partial^2\over\partial\phi^2} \nonumber
\EN
is the angular part of the Laplacian operator
\EQ
\label{eq_laplacian}
\nabla^2  =  {1\over r^2} {\partial \over \partial r}
r^2 {\partial \over \partial r}
+ {1\over r^2} L^2.   \nonumber
\EN
The  inversions in Eqs.(\ref{eq_rdotcurl}-\ref{eq_rdotcurl6}) 
can be efficiently  carried out numerically. 
%by inversion of  
%a tridiagonal matrix.
%
%~~~~~~~~~~~~~~~~~~~~~~~~~~~~~~~~~~~~~~~~~~~~~~~~~~~~~~~~~~~~~~~~~~~~~~
\subsection{The m-Modal Expansion}
%~~~~~~~~~~~~~~~~~~~~~~~~~~~~~~~~~~~~~~~~~~~~~~~~~~~~~~~~~~~~~~~~~~~~~~
We expand all functions $T, S, V, U$ in Eq.~(\ref{eq_tp}) into the
m-modes
\EQ
\label{eq_mmode}
T(r,\theta,\phi)=\sum_{m=0}^{N} T^m(r,\theta) e^{im\phi} +
{\rm cc, ...}
\EN
where $T^m(r,\theta)$ are complex functions and $N$ is the maximum order 
of the expansion.
When the sources ($\uu, \alpha, \eta$) are axisymmetric, i.e. do not
depend on the azimuth $\phi$,
Eqs.~(\ref{eq_tp}) decouple into a set of $2\, (N+1)$ equations.
\EQA
\label{eq_tpm}
\partial_t T^m &=&
R_\Omega V^m_{\Omega} + R_\alpha V^m_{\alpha}  + R_M V^m_{M}
   \nonumber \\
&+&\eta \nabla^2_m T^m_{\Omega}
+  \partial_r \eta(r) \frac{1}{r}\partial_r (rT^m),
\nonumber \\
\partial_t S^m &=&
R_\Omega U^m + R_\alpha U_{\alpha}^m + R_M U_{M}^m
\nonumber \\
&+&\eta \nabla^2_m S^m.
\ENA
where $\nabla^2_m$ is the 
 Laplacian operator (\ref{eq_laplacian}), in which 
$\partial^2/\partial^2\phi$ is replaced by $-m^2$.
Each of the m-modes
can therefore  be found from an independent set of equations.
The functions $U^m$, $V^m$,$U_{\alpha}^m $, $ V^m_{\alpha} $, $  U_{M}^m $,
and  $ V^m_{M}$,   are solutions of the m-transformed
equations (\ref{eq_rdotcurl}-\ref{eq_rdotcurl6}), such as:
\EQA \label{eq_moperators}
[-L^2]^m \;  U^m_{\alpha}(r,\theta)  \cdot e^{im\phi} =
\nonumber \\
= [\rrr\cdot \nab\times]^m \; \alpha\BB^m(r,\theta)
\cdot   e^{im\phi},
\ENA
\EQA \label{eq_moperators2}
[-L^2]^m \;  V^m_{\alpha}(r,\theta)  \cdot e^{im\phi} =
\nonumber \\
= [ \rrr\cdot  \nab\times \nab\times]^m \; \alpha\BB^m(r,\theta)
\cdot   e^{im\phi},
\ENA
for the potentials $U^m_{\alpha}$, $V^m_{\alpha}$. 
 Analogous expressions can be found  for the other potentials in Eqs.~(\ref{eq_rdotcurl}-\ref{eq_rdotcurl6}).
The operators 
$[ \rrr\cdot \nab\times]^m$, $[ \rrr\cdot  \nab\times
\nab\times]^m$
and $[-L^2]^m$
are  the m-transformed operators, where the
derivatives with respect to $\phi$ become
multiplications by the complex  factor $im$.

We solve the set of equations (\ref{eq_tpm}) numerically.
At each time step,
we solve Eqs.~(\ref{eq_moperators},\ref{eq_moperators2})
 and the
corresponding
equations for the $\uu\times\BB$ term 
for the auxiliary potentials 
\scriptsize{$V_{\Omega}^m,  U_{\Omega}^m,V_{M}^m,  U_{M}^m,  U^m_{\alpha},
V^m_{\alpha}$}\normalsize .
Equations (\ref{eq_tpm}) are then advanced in time with a third-order Runge
Kutta method. Spatial derivatives are second-order accurate in a regular
$[r,\theta]$ mesh.
The models below have the  resolution of 80x160 mesh points. 

%~~~~~~~~~~~~~~~~~~~~~~~~~~~~~~~~~~~~~~~~~~~~~~~~~~~~~~~~~~~~~~~~~~~~~~
\subsection{Coupling of the Modes}
%~~~~~~~~~~~~~~~~~~~~~~~~~~~~~~~~~~~~~~~~~~~~~~~~~~~~~~~~~~~~~~~~~~~~~~
When the assumption of axisymmetry for $\alpha$ is relaxed,  
the set of dynamo equations (\ref{eq_tpm}), for each $m$, are no longer 
independent.  
But it is still possible to use
Eqs.(\ref{eq_moperators},\ref{eq_moperators2}), where the product
$\alpha(r,\theta)\BB^m$ is replaced by the convolution
\EQ
\label{eq_alpBm}
(\alpha\BB)^m = \sum_{j=-\infty}^{\infty} \alpha^j(r,\theta)
\BB^{m-j}(r,\theta).
\EN
The coefficients with 
negative indices are equivalent to the complex conjugate of the coefficient with the positive indices.  

We investigate in detail the simplest situation of coupling of m = 0 and m =1 modes,  retaining only the lowest $m=0$ and $m=1$ terms in the expansion (\ref{eq_alpBm}).  The contribution to the $m=1$ mode thus becomes:
\EQ
\label{eq_alpB1}
(\alpha \BB)^1=\alpha^0 \BB^1 + \epsilon \alpha^1 \BB^0.
\EN
The non-axisymmetric part of $\alpha$ ($\alpha^1$) thus introduces a 
linear coupling between the $m=1$ and the $m=0$ modes. 
The amplitude of the resulting non-axisymmetric field depends on the
value of $\epsilon$. 
In the results displayed below, $\epsilon=1$ is used in model M1 and 
 $\epsilon=0.2$ in models M2 and M3. 
For an $\alpha\Omega$ dynamo, when $R_{\omega} \gg R_{\alpha}$, 
Eq.~(\ref{eq_alpB1}) means that the poloidal $m=1$ component
of the field is driven by the toroidal $m=0$ component.
The toroidal $m=0$, $m=1$ modes are related respectively 
to the poloidal $m=0$, $m=1$ modes 
by the differential rotation. 
%~~~~~~~~~~~~~~~~~~~~~~~~~~~~~~~~~~~~~~~~~~~~~~~~~~~~~~~~~~~~~~~~~~~~~~
\subsection{Boundary conditions}
%~~~~~~~~~~~~~~~~~~~~~~~~~~~~~~~~~~~~~~~~~~~~~~~~~~~~~~~~~~~~~~~~~~~~~~

%~~~~~~~~~~~~~~~~~~~~~~~~~~~~~~~~~~~~~~~~~~~~~~~~~~~~~~~~~~~~~~~~~~~~~~
%\subsubsection{Regularity Conditions on the Axis}
%~~~~~~~~~~~~~~~~~~~~~~~~~~~~~~~~~~~~~~~~~~~~~~~~~~~~~~~~~~~~~~~~~~~~~~
The conditions on the rotation axis $\theta =0$
are different for different values of $m$.  For even $m$,
any function maps into itself when $\phi \to \phi+\pi $ and  therefore
on the axis the derivative in the $\theta$-direction perpendicular to
the axis must vanish.  For odd $m$  every function subjected to this
mapping is multiplied by $-1$ and therefore must vanish on the axis
($\theta
=0) $ to satisfy continuity.  Hence,
\EQ
{\partial S \over \partial \theta} = {\partial T \over \partial \theta}=0,
\qquad  m = 0,
\EN
\EQ
S=T=0,
\qquad  m = 1.
\EN

We consider a spherical shell $r_{in} < R_{\odot} < 1$
with internal radius  $r_{in}=0.4
R_{\odot}$,
where we define our potentials $S$, $T$ to be zero.
The physical boundary between the core and the convection zone
is identified  by the sharp change 
in electrical resistivity at $0.69  R_{\odot}$.
The  computational domain extends further down,  to $0.4 R_{\odot}$. 
We allow enough time for the diffusion and differential rotation to expel
the flux from the low-conductivity region. 

On the outer boundary, $r=R$, we adopt the potential (vacuum) field
conditions:
\EQ
T=0 \; , \qquad
\frac{\partial S_{\ell}^{ m}}{\partial r}+\frac{(\ell+1)}{R}S_{\ell}^{m}=0,
\EN
where  $S_{\ell}^{ m}$ are the coefficients of  the decomposition 
into spherical harmonics of degree ${\ell}$ and order $m$ of the scalar
potential $S$ at the surface. 
%CHANGED
%The poloidal potential $S$ is obtained by solving the equation 
%$\nabla_{\rm H}^2S=-B_r$ near the boundary
%($\nabla_{\rm H}^2=\Drr+\Ddt$ is the horizontal Laplacian).
 For the numerical  scheme,  which is second order in the $r$-direction, 
 we only need to solve this equation at two r-layers  which are next to the boundary.
%~~~~~~~~~~~~~~~~~~~~~~~~~~~~~~~~~~~~~~~~~~~~~~~~~~~~~~~~~~~~~~~~~~~~~~
\subsection{Meridional Circulation}
%~~~~~~~~~~~~~~~~~~~~~~~~~~~~~~~~~~~~~~~~~~~~~~~~~~~~~~~~~~~~~~~~~~~~~~
Meridional circulation has long been observed on the surface of the Sun 
with a variety of techniques from sunspot tracking \cite{tuominen41} to 
Doppler and helioseiseismic measures \cite{LabonteHoward82,HathGONG96,Haber02}.
 
The patterns correspond to a single-cell large-scale flow proceeding from
the equator to the pole with an amplitude at the surface of about 
$20 \  {\rm m/s}$. Recent measurements indicate, however, 
that the flow might in fact have a more complex nature, even reversing at
times and showing a non-trivial vertical structure \cite{Haber02}. 

With current techniques, the upper $15\%$ of the solar surface can be
probed \cite{BraunFan98}. The return flow is not observed although some progress has been made in this direction \cite{giles97}.  In particular, the depth into the convection zone to which it penetrates is not known. Hathaway et al. (2003) examine the drift of the centroid of the sunspot area toward the equator in each hemisphere from 1874 to 2002 and find that these observations are consistent with
a meridional counterflow deep within the Sun as the primary driver of the migration toward the equator and the period associated with the sunspot cycle.

In the context of an axisymmetric dynamo, meridional circulation has been
extensively studied \cite{RobStix72, wsn91, ChouSchuDik95,Durney95,DikGil01, NandyCh02}. 
For the magnetic fields to be effectively transported by such a small amplitude 
flow, the electrical resistivity has to be low, a condition that is fulfilled below
the convection zone, where resistivity and turbulence dramatically decrease. 
Hence, one can expect that a deep flow, sinking below the convection zone,
will be more effective in acting upon the field \cite{NandyCh02}.  
However, it has not been established that such a flow can realistically exist. 
We will discuss two models, see Fig.~\ref{fig_merid}. 

%~~~~~~~~~~~~~~~~~~~~~~~~~~~~~~~~~~~~~~~~~~~~~~~~~~~~~~~~~~~~~~~~~~~~~~
\subsection{Initial Conditions}
%~~~~~~~~~~~~~~~~~~~~~~~~~~~~~~~~~~~~~~~~~~~~~~~~~~~~~~~~~~~~~~~~~~~~~~
We select an initial field of antisymmetric (dipolar) symmetry with respect
to the equator by choosing the following potentials for $m=0$ and $m=1$
modes:
\EQA
S(t=0)=S_1^0, \qquad S^1(t=0)=S_2^1,
\nonumber \\
T(t=0)=T_2^0, \qquad T^1(t=0)=T_1^1.
\ENA
Here, $S_l^m$ indicates a function proportional to the spherical harmonic of
order $l$ and degree $m$. 
No symmetry with respect to the equator is imposed on the time-dependent
solution, and the whole domain $\theta \in [0,\pi]$ is therefore included. 

The velocity field $\uu$ in the induction equation 
(\ref{eq_induction}) is given.
The induction equation is thus 
linear in the magnetic field $\BB$ and its solutions
are exponentially growing or decaying in time.
A stationary state is often achieved by introducing a quenching
of the $\alpha$-effect
when the field intensities exceed some level,  
for a discussion see for instance \cite{BlFie2002}.
We  do not prescribe here  any quenching mechanism because it 
would  introduce a further coupling between modes, making the interpretation of the
results more complex. It would also add an element of arbitrariness, which is not 
necessary for our purposes.  We will therefore be looking at
solutions that are close to marginal excitation. 
%~~~~~~~~~~~~~~~~~~~~~~~~~~~~~~~~~~~~~~~~~~~~~~~~~~~~~~~~~~~~~~~~~~~~~~
\section{Results}
%~~~~~~~~~~~~~~~~~~~~~~~~~~~~~~~~~~~~~~~~~~~~~~~~~~~~~~~~~~~~~~~~~~~~~~
We consider first the dynamo without meridional circulation ($\uu_M=0$).
In Table~\ref{tab_result} the  growth rate $\gamma$, the adimensional cycle
period $T$ and the ratio of the cycle period to the core rotation
period $ T/T_{rot}$ as explained in Section~\ref{s_model}, are given.
The three models considered are labelled as M1, M2, M3.
See also the Appendix for a more detailed description.

%~~~~~~~~~~~~~~~~~~~~~~~~~~~~~~~~~~~~~~~~~~~~~~~~~~~~~~~~~~~~~~~~~~~~~~
\subsection{Growth rates and cycle periods}
\begin{figure}
%\figurenum{<text>}
\epsscale{1.}
\plotone{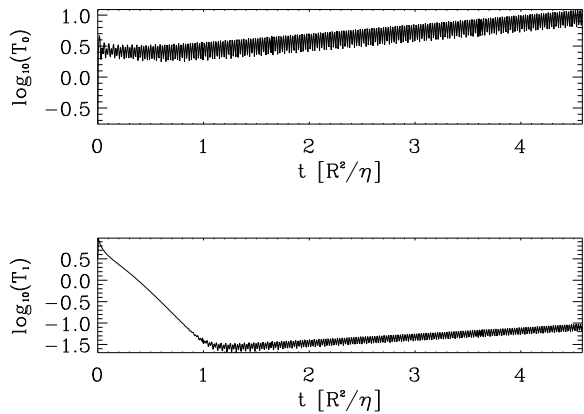}
\caption{
Long-term time evolution of the amplitude of the toroidal field potentials
for model M1. 
Top panel is $T_0$ and lower panel is $T_0$. 
The diffusion time is taken as the time unit.
\label{fig_evol}
}
\end{figure}

%~~~~~~~~~~~~~~~~~~~~~~~~~~~~~~~~~~~~~~~~~~~~~~~~~~~~~~~~~~~~~~~~~~~~~~
The time evolution of the toroidal field potential for the  M1 case
is shown in  Fig.~\ref{fig_evol}. 
A steady exponential growth is reached after one diffusion time. 
We follow the run for approximately five diffusion times.  
However, already 
 after $t\sim1 $, the
solution appears to be stable. 
As seen in Fig.~\ref{fig_evol}, 
the solution is oscillating. Each cycle lasts 179 rotations of the body being
modelled. 
The ratio of the cycle period  to the rotation rate, ($T/T_{rot}$)
is independent of the units adopted.  This ratio discriminates between
solutions that have a solar-like period and those which do not.
For model M1, the ratio $T/T_{rot}$ is 179 and is close to the
solar case of $149$. 

Model M3, not shown, behaves similarly to model M1. 
It has a comparable oscillation period. 
The $\alpha$-effect in these two models 
 have  a broad radial distribution within the convection zone which 
overlaps with the shear layer, see Fig.~\ref{fig_alp}. 

In contrast, model's  M2 behaviour  
is markedly different than the previous two, especially regarding to  the
oscillation period $T$, which is ten times smaller than the cases M1 and
M3. $T/T_{rot}$ of $20.2$ shows that M2 does not represent the solar case
correctly. 
M2 is characterised by the  
$\alpha$-effect which has no overlap with the shear layer and this is probably
the cause of its failure.

%
%~~~~~~~~~~~~~~~~~~~~~~~~~~~~~~~~~~~~~~~~~~~~~~~~~~~~~~~~~~~~~~~~~~~~~~
\subsection{Phase relations between the axisymmetric and non-axisymmetric
modes}
\label{s_phase}
\begin{figure}
%\figurenum{<text>}
\epsscale{1.}
\plotone{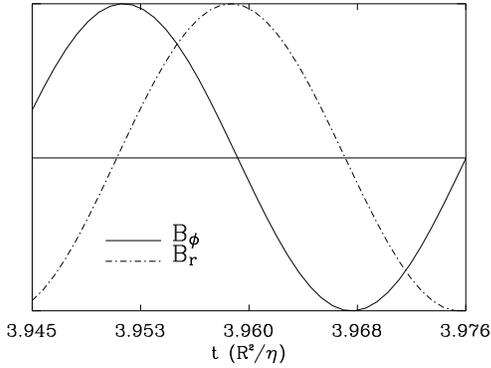}
\caption{
Phase relation between magnetic field components in model M3. 
The time evolution during a cycle of magnetic field components is plotted. 
The solid line shows the intensity of the azimuthal component of the
total magnetic field, $\BB^0_{\varphi}+ \BB^1_{\varphi}$,
in the meridional plane $\varphi=0$,
calculated 
 at  the 
tachocline, $r=0.69 R_{\odot}$, at $30^{\circ}$ latitude. 
The dot-dashed line shows the intensity of the 
surface radial field component $\BB^0_{r}$ close to the north pole, at $80^{\circ}$ latitude. The phase shift between the fields is $\pi/2$. 
Amplitudes of both toroidal and radial fields are normalised to unity 
for convenience.
\label{fig_phase_fields}
}
\end{figure}

%~~~~~~~~~~~~~~~~~~~~~~~~~~~~~~~~~~~~~~~~~~~~~~~~~~~~~~~~~~~~~~~~~~~~~~
In Fig.~\ref{fig_phase_fields} we show the time evolution of the radial and
azimuthal components of the magnetic field during one cycle in model M3. The
axisymmetric dipolar-type field at the pole reverses its sign when the (total)
azimuthal field close to the tachocline is maximal.
In other words, there is a phase shift of $\pi/2$ between these two field components. 
We evaluate the toroidal fields at $30^{\circ}$ of latitude, where the non-axisymmetric field is concentrated. For the plot of the $m=1$ component, which has the $\cos \varphi$ dependence, we select the meridional plane defined by $\varphi=0$, where the field is maximum. A similar relation is observed on the Sun: the axisymmetric dipole reverses close to the solar maximum. 
Analogous calculations for models M1 and M2 yield $\pi/6$ and $-3/4\pi$,
respectively,
see Table~\ref{tab_phase}. 
We may therefore conclude that the phase relation between the field components  is 
dependent on the radial distribution of the $\alpha$-effect. This can help to select the models according to the phase relations between the field components, as originally indicated by Stix (1976).

\begin{figure}
%\figurenum{<text>}
\epsscale{1.}
\plotone{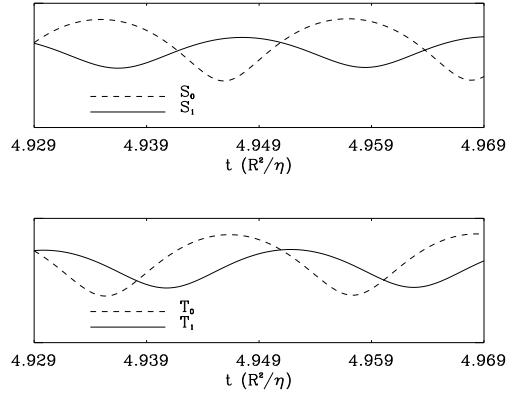}
\caption{
Phase relations between poloidal and toroidal components of the
axisymmetric and non-axisymmetric magnetic fields for model M1. 
The integrated square
moduli of the scalar potentials S,T, over the whole domain are plotted
versus time. Values are not to scale.
 $T_0$ and  $S_0$ are the toroidal and poloidal potentials for the $m =
 0$ mode,
$T_1$ and  $P_1$ are the toroidal and poloidal potentials for the $m=1$
mode.
The $S_1$ mode is coupled  to the  $T_0$ mode
 (see Eq.~\ref{eq_alpB1}).
The observed phase relations for the Sun's field components are shown in
Fig.~\ref{fig_phase_obs} for comparison. 
\label{fig_phase}
}
\end{figure}

\begin{figure}
%\figurenum{<text>}
\epsscale{1.}
\plotone{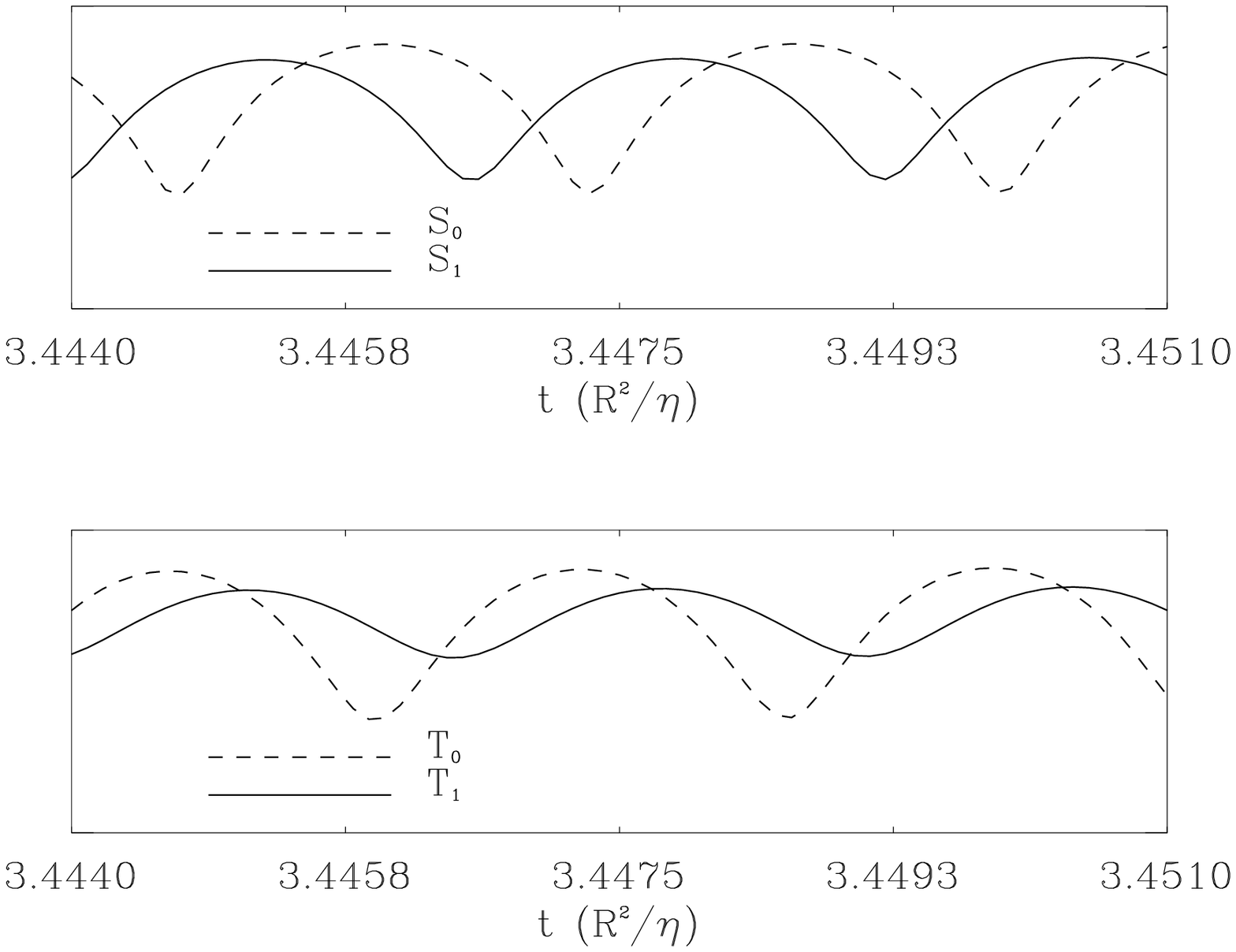}
\caption{
Same as in Fig.~\ref{fig_phase} for model M2. 
\label{fig_phaseM2}
}
\end{figure}

\begin{figure}
%\figurenum{<text>}
\epsscale{1.}
\plotone{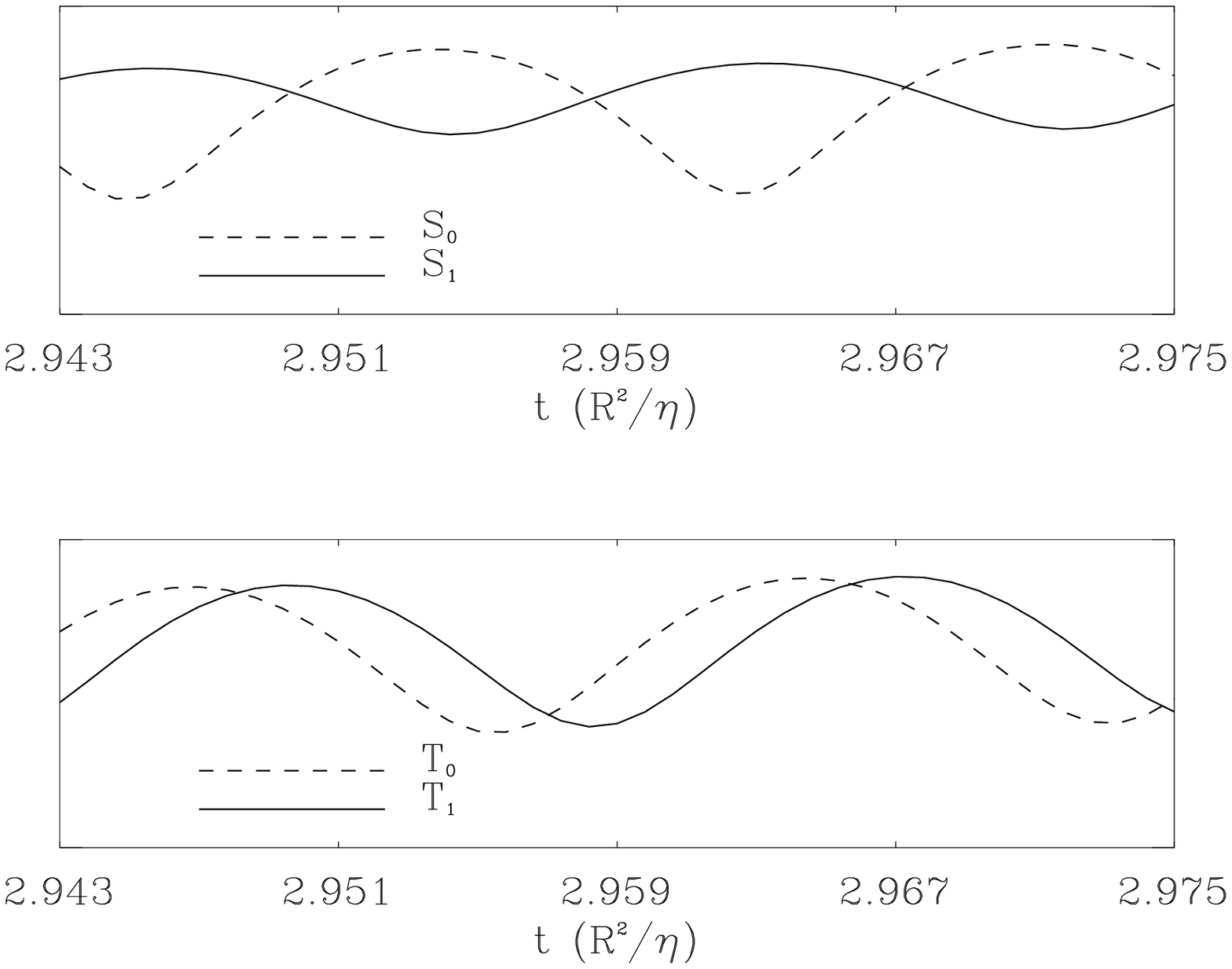}
\caption{
Same as in Fig.~\ref{fig_phase} for model M3. 
\label{fig_phaseM3}
}
\end{figure}

\begin{figure}
%\figurenum{<text>}
\epsscale{1.}
\plotone{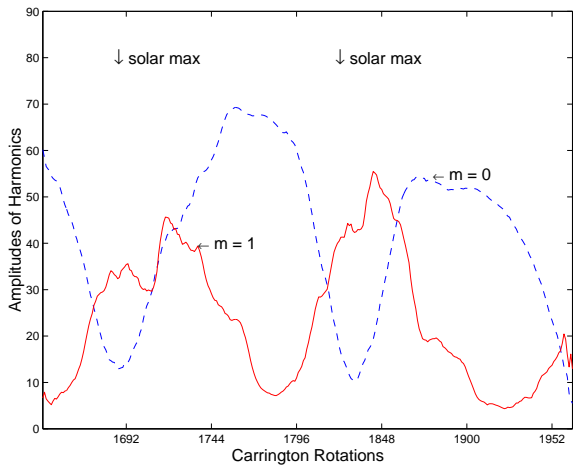}
\caption{
Observed phase relations between poloidal components of the
axisymmetric (dashed line)  and non-axisymmetric 
(solid line) magnetic field \cite{ruzmaikin01b}.
\label{fig_phase_obs}
}
\end{figure}

Another phase relation that can be observed 
is the one between the surface axisymmetric and non-axisymmetric poloidal 
fields is shown in Figure~\ref{fig_phase_obs}  \cite{ruzmaikin01b}. 
Note that, according to their definitions, the poloidal potentials have the
same phase as the poloidal fields. For the purpose of determining
the phase difference between the modes of the poloidal field, we can therefore use the potentials, see Figs.~\ref{fig_phase}-\ref{fig_phaseM3}. 

The time dependence of the $m=0$ and $m=1$ poloidal potentials
is shown in  Figs.~\ref{fig_phase}-\ref{fig_phaseM3}. The phase relations can be read from the plot and are listed in Table~\ref{tab_phase}. 

%~~~~~~~~~~~~~~~~~~~~~~~~~~~~~~~~~~~~~~~~~~~~~~~~~~~~~~~~~~~~~~~~~~~~~~
\subsection{Spatial distribution of the fields. Localisation of the non-axisymmetric modes}
\label{s_distr}
\begin{figure}
%\figurenum{<text>}
\epsscale{1.}
\plotone{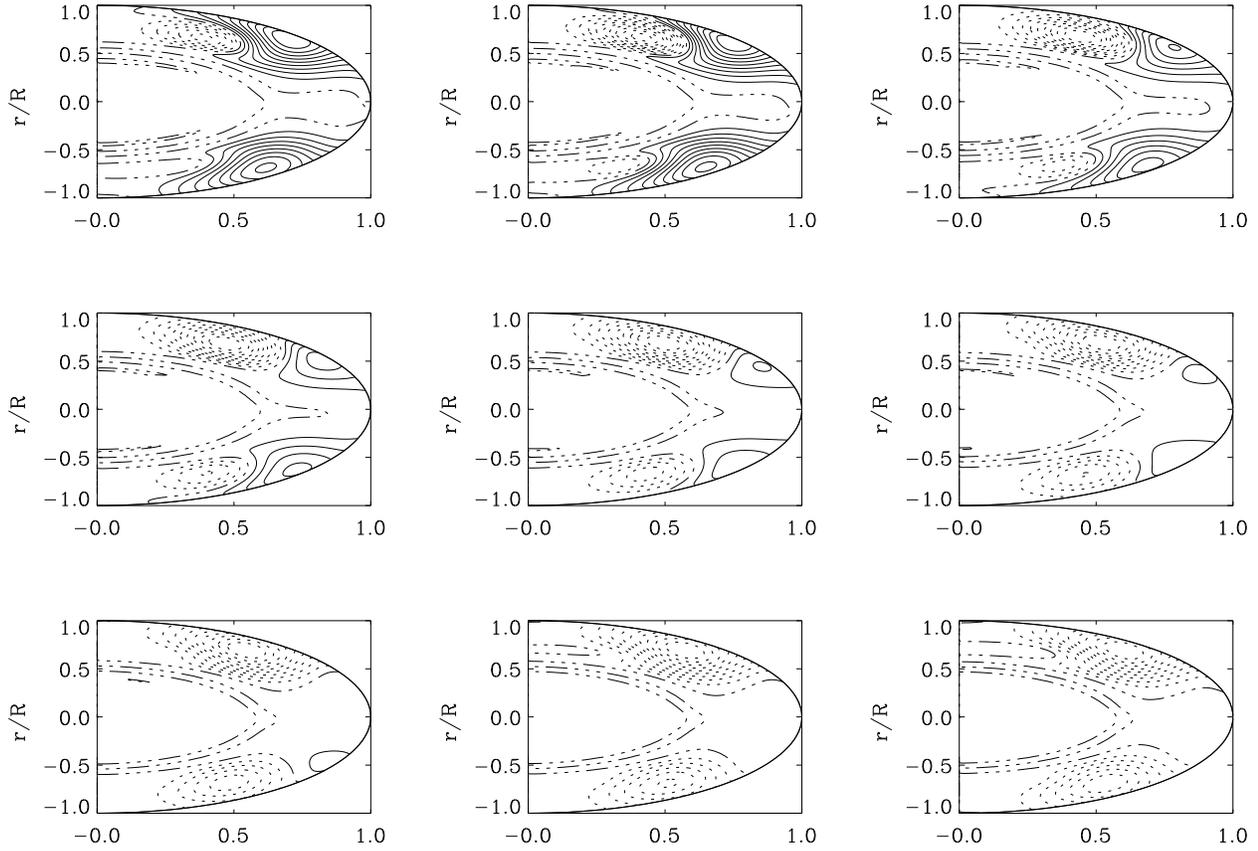}
%\plottwo{<epsfile>}{<epsfile>}
\caption{
Distribution of  the axisymmetric $(m=0)$ poloidal field over half a
cycle. Time proceeds from left to right and from top to bottom and panels are equally
spaced in time. Poloidal field lines are shown. Solid lines have a counter-clock wise
direction. Dotted lines mean opposite direction  
(model M1). 
\label{fig_polo}
}
\end{figure}

\begin{figure}
%\figurenum{<text>}
\epsscale{1.0}
%\plotone{f11.ps}
\plottwo{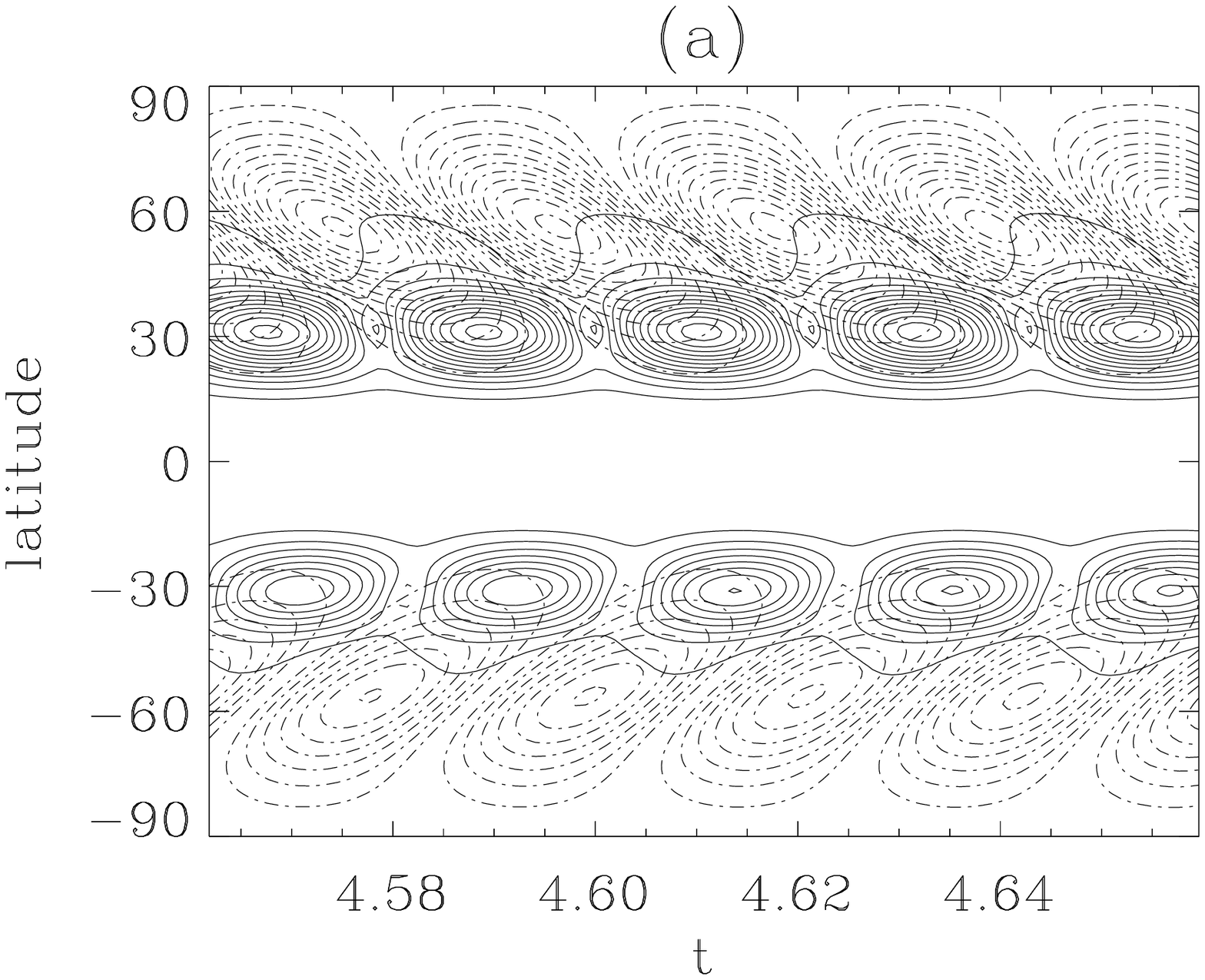}{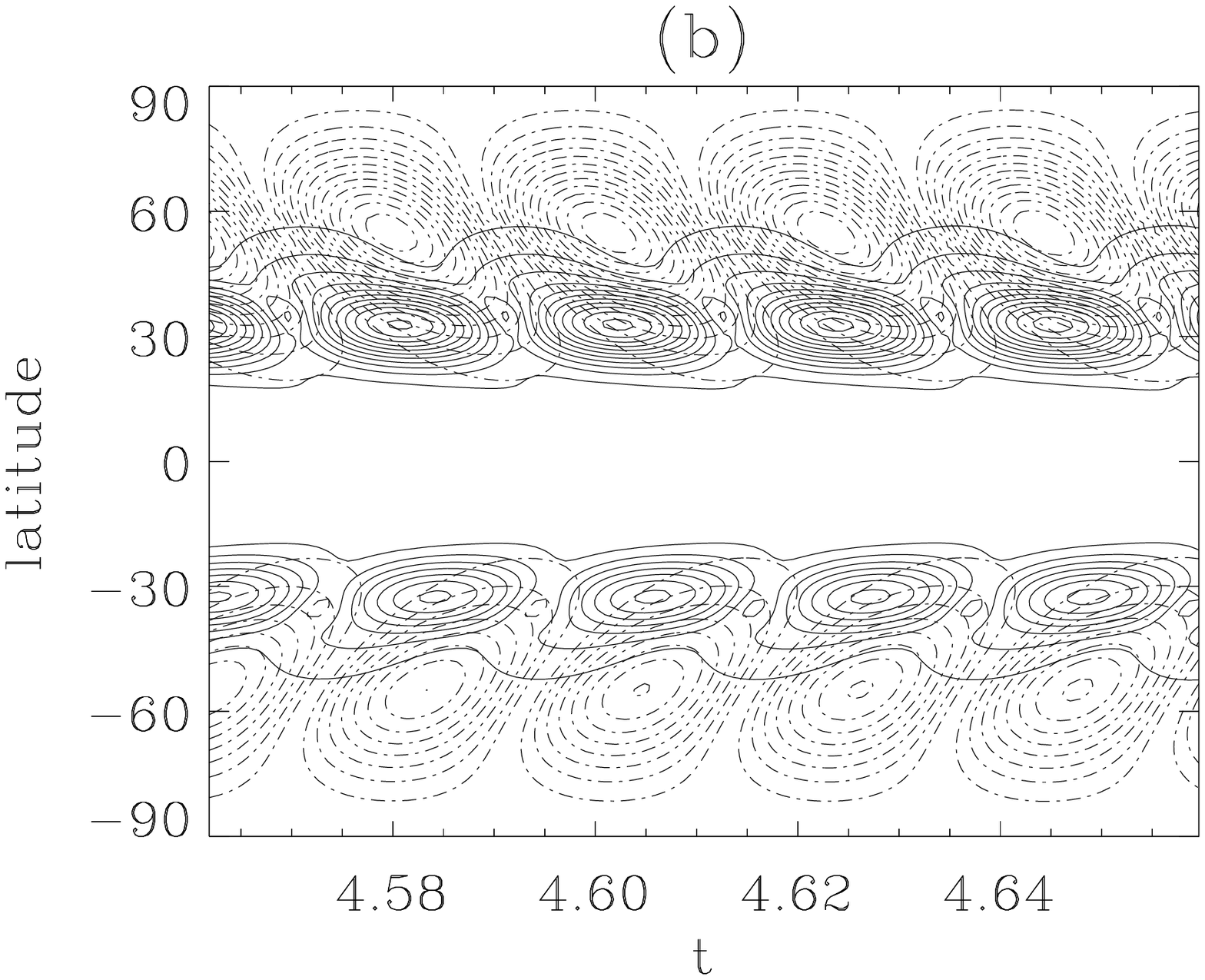}
\caption{
Butterfly diagram for the $m=0$ (dot-dashed contours) 
and $m=1$ (solid contours) modes of the toroidal
magnetic field at two different depths: above the shear layer,  
$0.74 R_{\odot}$ (a),  and
in the convection zone, at $0.85 R_{\odot}$ (b). 
Model M1 is displayed here. 
The diagram shows the latitudinal distribution of the 
modulus of the toroidal magnetic field as
a function of time, during  five half-cycles. The time is measured in unit
of the diffusion time.
\label{fig_butfly}
}
\end{figure}

\begin{figure}
%\figurenum{<text>}
\epsscale{1.}
%\epsscale{.8}  % APJ
	\plotone{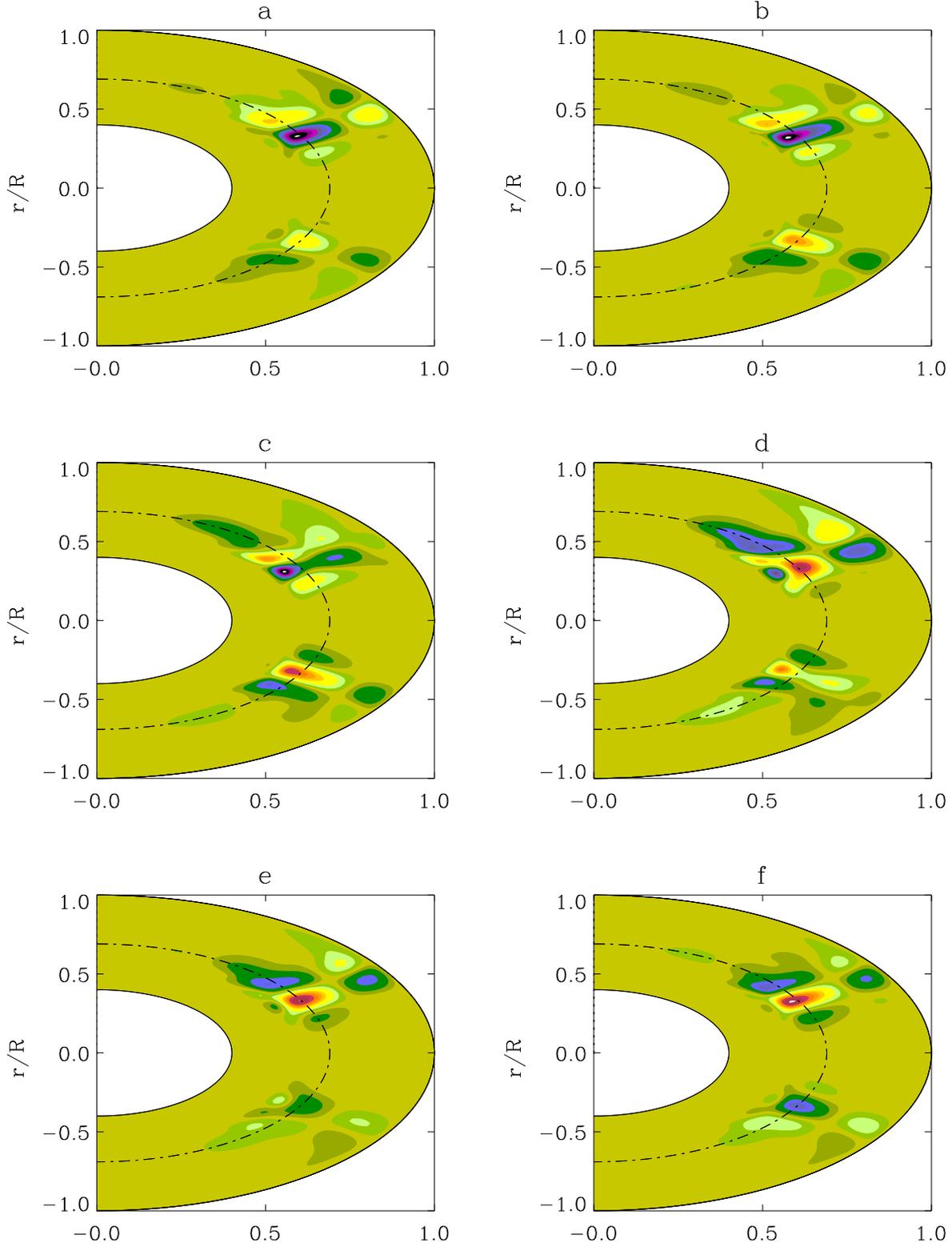}
	%\plottwo{<epsfile>}{<epsfile>}
	\caption{
	Distribution of the  
	non-axisymmetric $(m=1)$ toroidal field   through half a
	cycle,  until the field has changed its polarity. 
	Model M1.  The contour plot show the intensity of the azimuthal component of
	the non-axisymmetric field $\BB^1_{\varphi}$. 
	The dashed line marks the centre of the tachocline, at $r=0.69 R_{\odot}$, 
	which is also the location of the turbulent resistivity drop.
	\label{fig_toro_t}
	}
	\end{figure}

%~~~~~~~~~~~~~~~~~~~~~~~~~~~~~~~~~~~~~~~~~~~~~~~~~~~~~~~~~~~~~~~~~~~~~~
The poloidal field is found to be predominantly
antisymmetric (odd) with respect to the equator 
(dipolar-type) in all three models. 
However, this symmetry is not pure and a symmetric
component is present as well, see Fig.~\ref{fig_polo},\ref{fig_butfly}. 
We recall that we do not impose an equatorial symmetry conditions
on the solution and solve the equations for the whole domain  $\theta \in [0,\pi]$. The north-south symmetry is broken because we couple the axisymmetric mode, which is odd across the equator (like polar dipole), with the non-axisymmetric mode, which is even across the equator (like equatorial dipole). This north-south asymmetry does not disagree with observations.
Fig.~\ref{fig_toro_t} shows the time evolution of the signed azimuthal magnetic
field for  the m = 1 mode on the meridional plane $\phi=0$ during half of a cycle. 
In contrast to the axisymmetric mode, the non-axisymmetric field is highly
concentrated in the lower part of the tachocline and, to lesser extent,
near the surface.  It is also localised around \(30^\circ\) latitude, see
also Table~(\ref{tab_distr}).

\begin{figure}
	%\figurenum{<text>}
	\epsscale{.6}
	%\epsscale{1.}
	\plotone{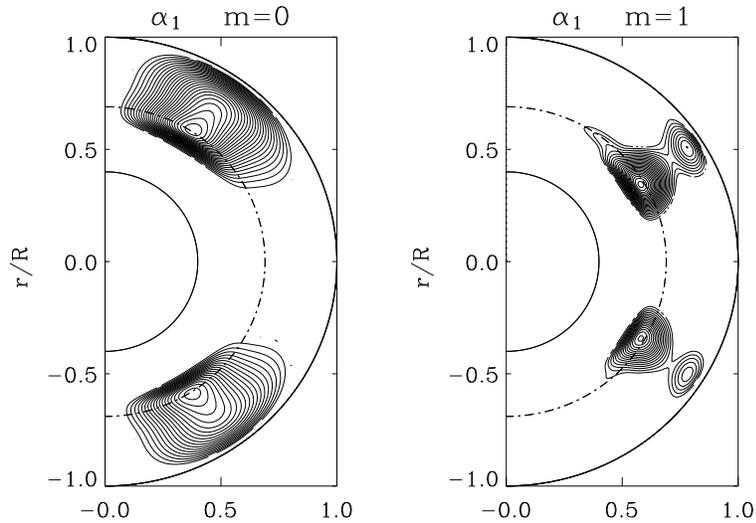}
	%\plottwo{<epsfile>}{<epsfile>}
	\caption{
	Distribution of   the axisymmetric $(m=0)$ and
	non-axisymmetric $(m=1)$ toroidal fields inside the convection zone in model
	M1. 
	The modulus of the azimuthal component of the magnetic field is shown. 
	Field distribution is integrated over time during a cycle. 
	Contour lines concentrate into region where the field  is more intense and
	resides for longer times.
	For the $m=1$ component, the modulus squared of the field
	is averaged over the azimuthal direction $\varphi$.
	The dashed line marks the centre of the tachocline, at $r=0.69 R_{\odot}$
	which is also the location of the turbulent resistivity drop.
	\label{fig_toro_m0m1-1}
	}
	\end{figure}

\begin{figure}
	%\figurenum{<text>}
	%\epsscale{1.}
	\epsscale{.6}
	\plotone{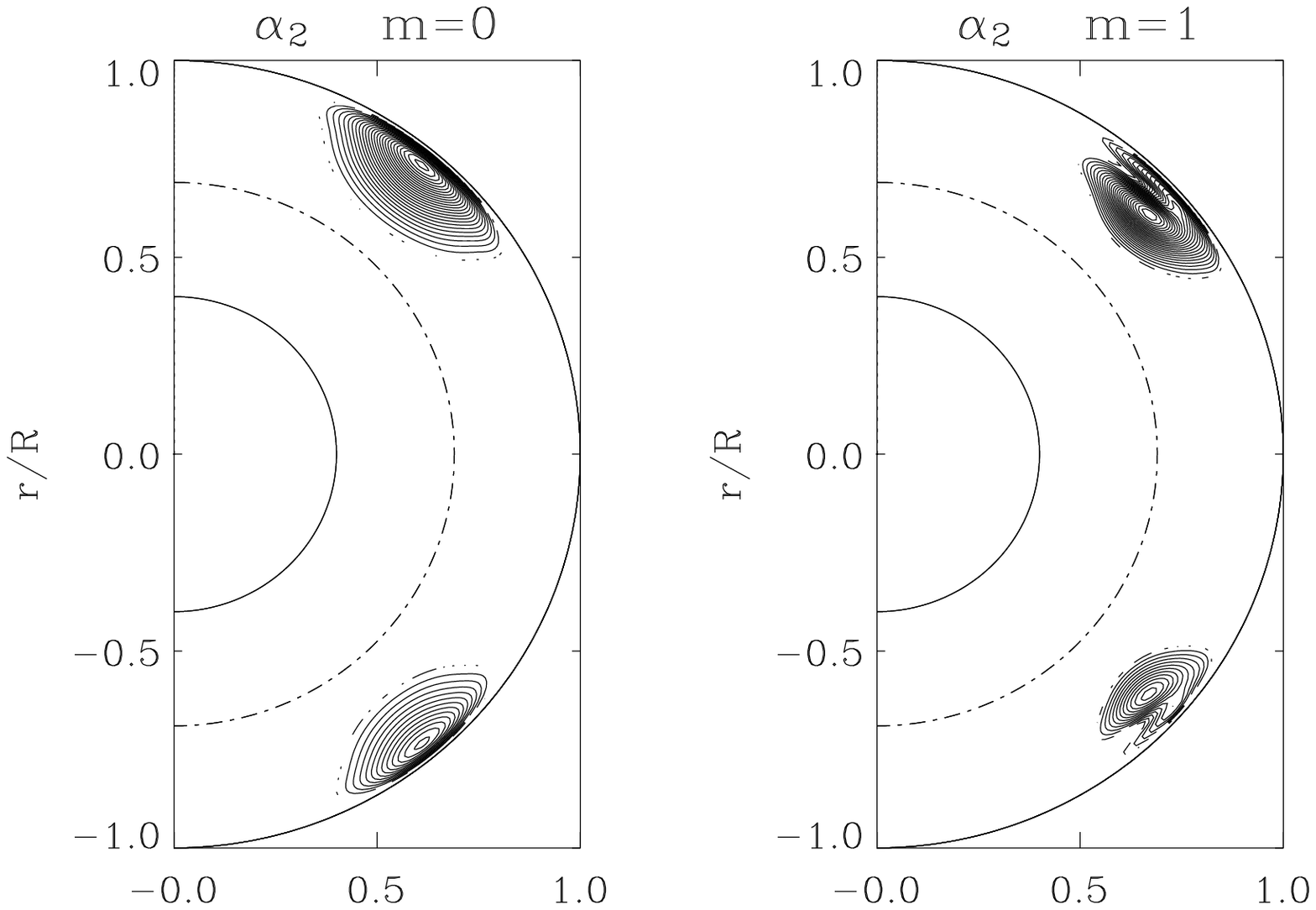}
	%\plottwo{<epsfile>}{<epsfile>}
	\caption{
	Same as in Fig.~\ref{fig_toro_m0m1-1} for model M2. 
	\label{fig_toro_m0m1-2}
	}
	\end{figure}

\begin{figure}
	%\figurenum{<text>}
	\epsscale{.6}
	\plotone{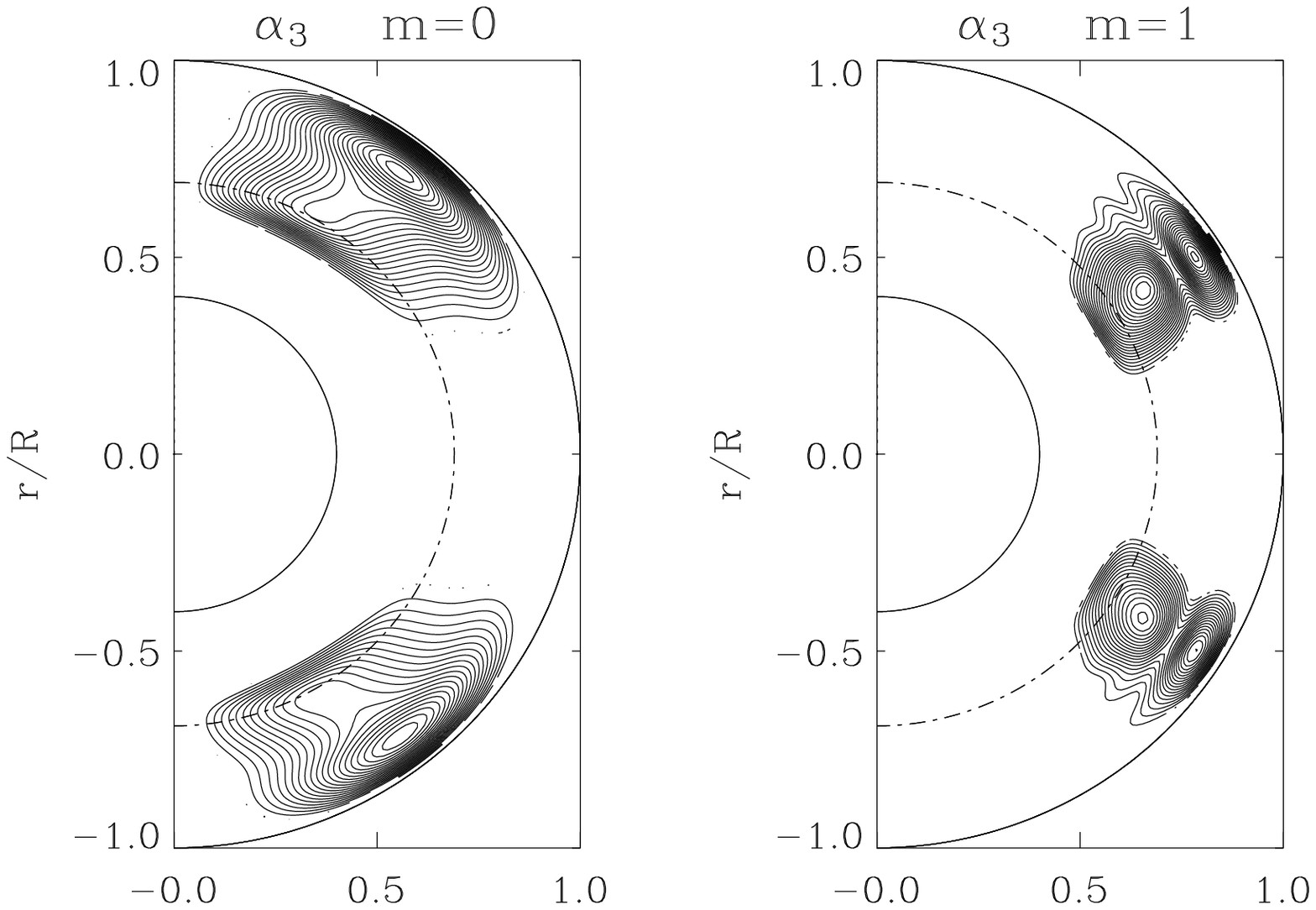}
	%\plottwo{<epsfile>}{<epsfile>}
	\caption{
	Same as in Fig.~\ref{fig_toro_m0m1-1} for model M3. 
	\label{fig_toro_m0m1-3}
	}
	\end{figure}

In order to show the robustness of the spatial intensity distribution of the fields
throughout a cycle we display the distribution of the absolute values of the 
the axisymmetric (m =0) and non-axisymmetric (m = 1) toroidal fields inside the convection zone, 
integrated over a solar cycle
(Figs.~\ref{fig_toro_m0m1-1},\ref{fig_toro_m0m1-2},\ref{fig_toro_m0m1-3}).
In Table~(\ref{tab_distr}) we show the locations of the maxima of the
toroidal $m=0$ and $m=1$ components, $r_M^0$, $\theta_M^0$, $r_M^1$ and
$\theta_M^1$.

The physical reason for these localisations can be understood.   As it
is well known from MHD theory\cite{moffatt78,KR80}, the differential
rotation affects the axisymmetric and non-axisymmetric fields in 
different ways.  Thus, an axisymmetric poloidal field
in a differentially rotating region is converted into
an axisymmetric toroidal field. But a non-axisymmetric field is rapidly
(within a few rotations) excluded from the strongly  differentially rotating
high resistivity region. This happens because the rotation tends to 
twist the field
lines perpendicular to the axis of rotation in such a way that any
two adjacent lines would have an opposite direction. This leads to
effective reconnection of these field lines, excluding the field from the 
differentially rotating region even in the case of a very
small electrical resistivity.  Hence, differential rotation
is a mechanism that tends to destroy any deviation from axial symmetry.
In the solar convection zone, the field survives only in the regions
where the differential rotation is weak (around \(30^\circ\)) and where
the resistivity is small, i.e. in the lower part of the tachocline.

The localisation of non-axisymmetric magnetic field suggests a simple
explanation of why active regions emerging on the solar surface have
a tendency to cluster on preferred longitudes. Active regions, such as
sunspots, are the result of buoyancy of bipolar magnetic loops (flux ropes)
through the solar surface, \cite{Parker79}.  The flux ropes formed at the
base of the convection zone emerge when their magnetic field exceeds the
threshold for buoyancy.  A non-axisymmetric enhancement of the underlying
magnetic field at that location results in the clustering of active regions
as shown in a heuristic model (Ruzmaikin 1998,2001).
%\cite{ruzmaikin98,ruzmaikin01a}. 
The \(m =1\) mode of the toroidal
magnetic field superimposed on the axisymmetric mode produces a localised
field maximum (``hump'') 
near the maximum of \(\sin\phi\).  The humps, produced by non-axisymmetric
fields, are unstable when their field strength reaches \(4\times10^4\)
G \cite{caligari95}.  Because the growth of 
the underlying non-axisymmetric mode
is affected by the continuous stretching by differential rotation, 
the mode is after some time destroyed.
The stretching time scale of several months is in agreement
 with the observed life-time of the clusters
 \cite{gaizauskas83}.

%~~~~~~~~~~~~~~~~~~~~~~~~~~~~~~~~~~~~~~~~~~~~~~~~~~~~~~~~~~~~~~~~~~~~~~
\subsection{Rotation Rate and Preferred Longitudes}
\label{s_rot}
%~~~~~~~~~~~~~~~~~~~~~~~~~~~~~~~~~~~~~~~~~~~~~~~~~~~~~~~~~~~~~~~~~~~~~~
As we mentioned in the Introduction,  a persistent, periodic magnetic pattern in the solar wind has been observed, with a fixed rotation rate \cite{neugebauer00}. A period consistent with this rate was also found from the surface fields \cite{ruzmaikin01b,henney03}. If preferred longitudes are associated with the non-axisymmetric modes of the dynamo, then the observed pattern would originate in
the poloidal non-axisymmetric component of the field. 
   
\begin{figure}
	%\figurenum{<text>}
	\epsscale{.6}
	%\plotone{./fig/omega_prof.eps}
	\plotone{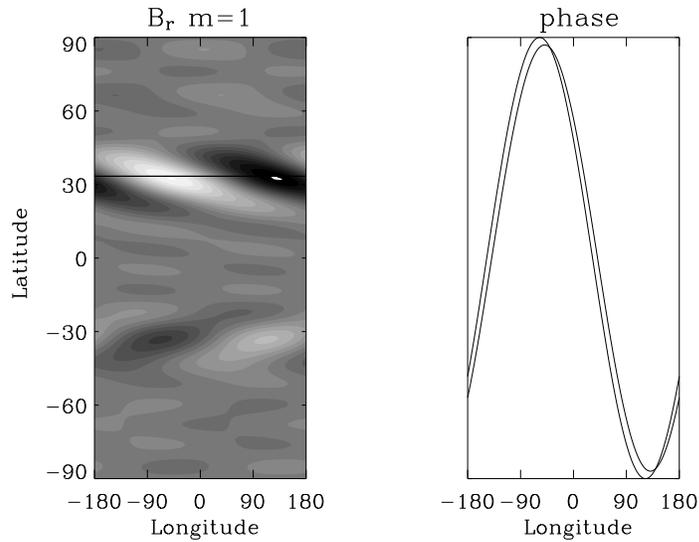}
	%\plottwo{<epsfile>}{<epsfile>}
	\caption{
	Left panel: Near-surface distribution of the radial component of the 
	non-axisymmetric magnetic field at fixed time.
	White indicates outwardly directed field, black inwardly. 
	Right panel: amplitude of the radial field versus longitude at the latitudinal
	cut indicated by the line in left panel. 
	Two curves are shown, obtained at different times during a cycle. 
	The phase difference between the maxima divided by the time lag between
	the two profiles returns the rotation rate of the $m=1$ mode. 
	\label{fig_magn_m1+phase}
	}
	\end{figure}

The longitudinal distribution of the non-axisymmetric modes of the dynamo has the form of a wave propagating along the azimuth \(\phi\) with a  rate  determined by the solution of the
dynamo equations. We determine that rate, $\Omega^1$, by following the time evolution of the surface poloidal (radial) $m=1$ mode, 
Fig.~\ref{fig_magn_m1+phase}. 
The results are given 
%in Table~\ref{tab_distr} for the three models
in Table~\ref{tab_result} for the three models
considered.
We represent these rates as horizontal dashed lines, along with the 
the radial distribution of the rotation rate, in Fig.~\ref{fig_rotOurs}.

The rotation  rate of the non-axisymmetric $m=1$ mode is
comparable to that of the surface rotation at $30^{\circ}$
latitude for the M1 and  M3 models. This is also the rotation rate of the
core. 
For the M2 model, a slower rate is found. 
Recent Ulysses observations of the surface magnetic  fields during Ulysses'
fast scan around solar maximum, show a rotation
rate of the non-axisymmetric dipole of the Sun close to the core rotation
rate, see \cite{JonesBal03}.

%~~~~~~~~~~~~~~~~~~~~~~~~~~~~~~~~~~~~~~~~~~~~~~~~~~~~~~~~~~~~~~~~~~~~~~
\subsection{Meridional Circulation}
%~~~~~~~~~~~~~~~~~~~~~~~~~~~~~~~~~~~~~~~~~~~~~~~~~~~~~~~~~~~~~~~~~~~~~~
The main goal of this paper is to study the excitation of the 
non-axisymmetric mode  under otherwise simple conditions. 
For this purpose we did not include the meridional circulation in the
models described above. Meridional circulation may play an important
role in the overall dynamo, as indicated by other studies
\cite{RobStix72, wsn91, ChouSchuDik95,Durney95,DikGil01, NandyCh02}.
Here we report only on a preliminary analysis of two simple cases of meridional
circulation for the M1 distribution of $\alpha$.  
The cases are (see Appendix): (1) shallow and (2) deep
meridional circulation, see Fig.~\ref{fig_merid}. 
In order to make the meridional
transport more effective compared to diffusion in these calculations  
we decrease the ratio of the core resistivity to the resistivity in the
convection zone to 
$1/200$.

\begin{figure}
	%\figurenum{<text>}
	\epsscale{.6}
	\plotone{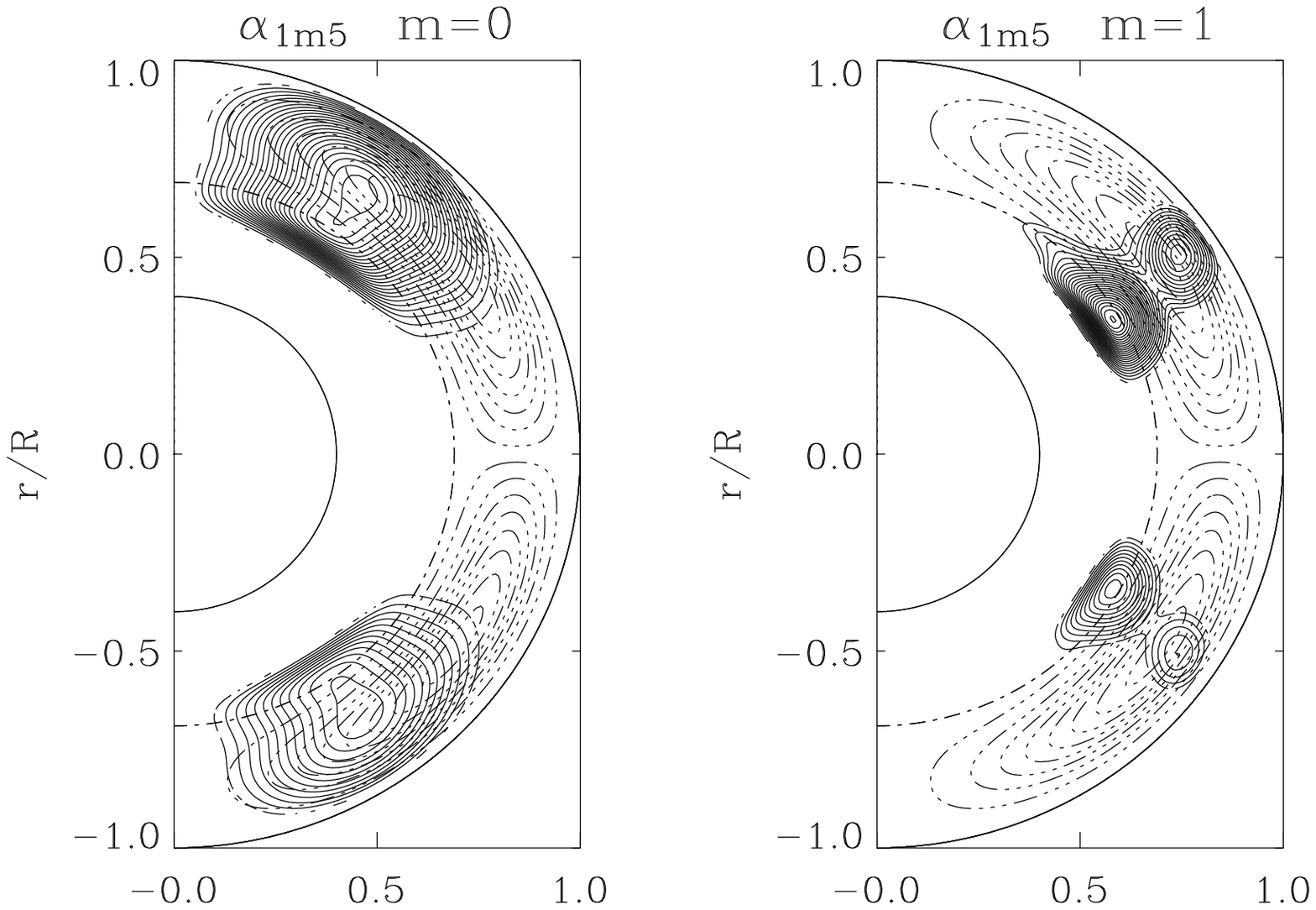}
	%\plottwo{<epsfile>}{<epsfile>}
	\caption{
	Same as in Fig.~\ref{fig_toro_m0m1-1} for model  
	with shallow meridional circulation. 
	The flow lines are superimposed as dot-dashed lines. 
	\label{fig_toro_m0m1-1ms}
	}
	\end{figure}

\begin{figure}
	%\figurenum{<text>}
	\epsscale{.6}
	\plotone{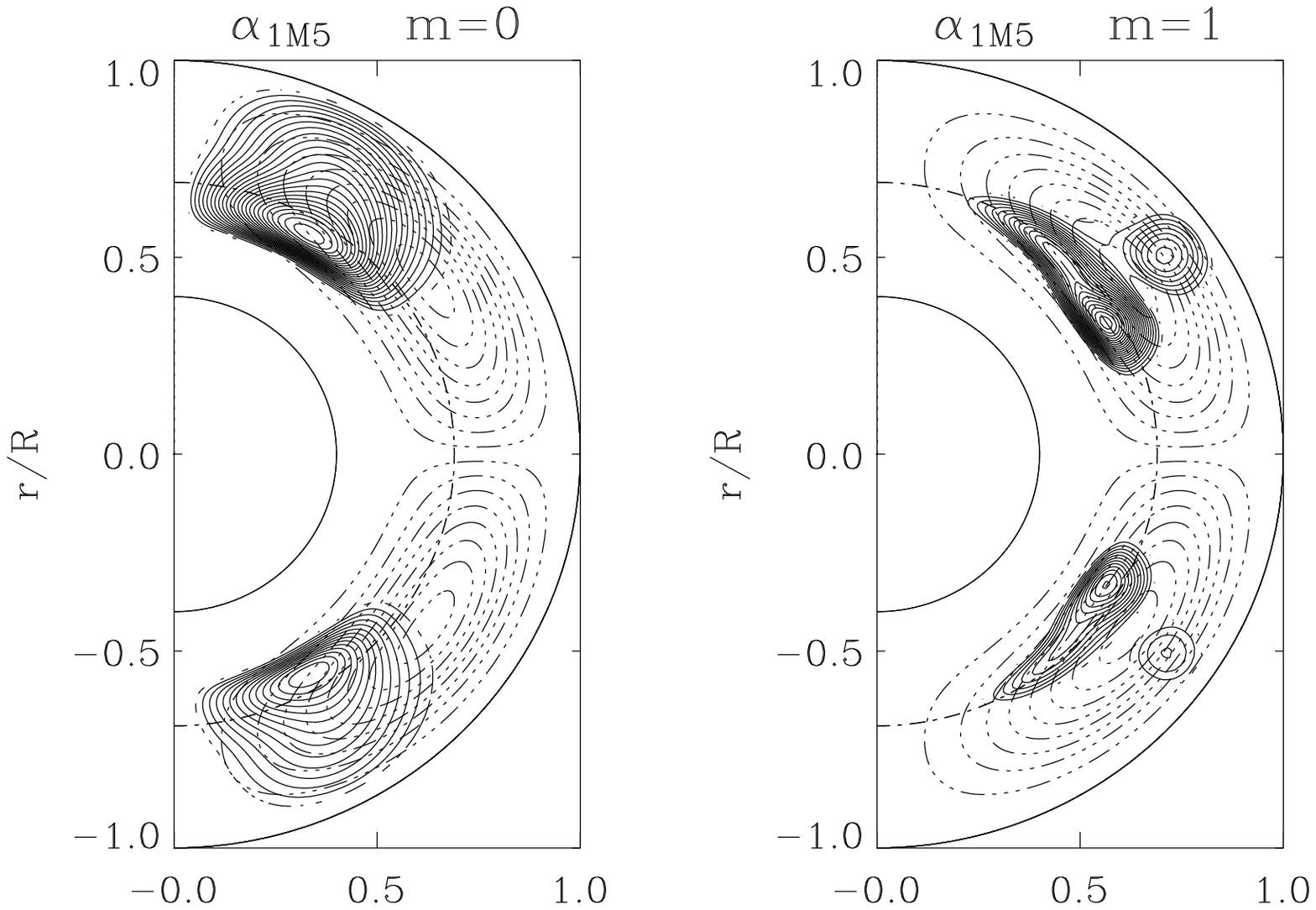}
	%\plottwo{<epsfile>}{<epsfile>}
	\caption{
	Same as in Fig.~\ref{fig_toro_m0m1-1} for model  
	with deep meridional circulation. 
	The flow lines are superimposed as dot-dashed lines. 
	\label{fig_toro_m0m1-1md}
	}
	\end{figure}

The resulting field distributions two cases are displayed in
Figs.~\ref{fig_toro_m0m1-1ms},\ref{fig_toro_m0m1-1md}. 
Comparing these distributions 
with the case without meridional circulation,
Fig.~\ref{fig_toro_m0m1-1}, we see that the field tends to be  depleted 
in the upper part of the convection zone and intensified and stretched in the
tachocline. This is especially true for the case with deep meridional
circulation. 

The field is pushed out of the 
convection zone into the tachocline, apparently under 
the action of the radial component of the circulation. 
The flow, however, 
does not seem to be
effective enough in the latitudinal direction 
to significantly change the toroidal field distribution, although
there is a shift of its maximum below the 
$30^\circ$ line (see Table~\ref{tab_merid}).

A more significant change occurs in the cycle period of the solutions,
which nearly doubles with an amplitude of the flow 
$|\uu_M=5|$ in comparison with the case $|\uu_M|=0$. 

Moreover, with $|\uu_M > 5|$, the symmetry of the solution changes from dipolar
(antisymmetric with respect to the Equator), to quadrupolar (symmetric). 
For $\uu_M=10$ a steady non-oscillatory solution takes
over.
%~~~~~~~~~~~~~~~~~~~~~~~~~~~~~~~~~~~~~~~~~~~~~~~~~~~~~~~~~~~~~~~~~~~~~~
\section*{Discussion}
%~~~~~~~~~~~~~~~~~~~~~~~~~~~~~~~~~~~~~~~~~~~~~~~~~~~~~~~~~~~~~~~~~~~~~~
Solar observations have shown the importance of the non-axisymmetric 
component of the mean magnetic fields 
in the formation and evolution of  solar activity. 
Solar dynamo models have so far failed to address the non-axisymmetric
features of the solar cycle, most notably the clustering of magnetic activity
and preferred longitudes. 

We have developed a new kinematic non-axisymmetric 
mean-field dynamo model of the Sun in spherical geometry, which
incorporates the solar rotation as reconstructed from helioseismic data
and a model for the meridional circulation. We decompose the magnetic
field into its toroidal and poloidal components. This geometry allows
waves to propagate azimuthally at different rates for different modes,
contrast to a periodic box where only one rate can be allowed. 
We determine the rotation rate of the first non-axisymmetric mode. 

We have considered for simplicity only the lowest azimuthal modes, $m=0$ and
$m=1$. By the action of a non-axisymmetric $\alpha$-effect, the modes are  coupled
together, and the $m=1$ mode shares the same cyclic behaviour  as the axisymmetric mode $m=0$. 

We have examined three different radial distributions of the mean
helicity ($\alpha$-effect), two with $\alpha$ operating near the
tachocline and one only near the surface. 
When the meridional flow is neglected, we found that all produced a
cycle with alternating  $m=0$ and $m=1$ modes. 
The two models with $\alpha$ near the tachocline both give the ratio
$T/T_{rot}$ of the cycle duration to the rotation rate 
within $20\%$ of the observed value of $147$, whereas the case where
$\alpha$ was concentrated near the surface failed to give agreement
indicating that this was not a valid solar model. 

The radial profile of $\alpha$ strongly 
influences the phase relations between the alternating axisymmetric
dipole and the toroidal field intensity at the base of the convection
zone.

We determined the rotation rate of the near-surface non-axisymmetric 
radial field in the three models. 
The two models with $\alpha$ close to the tachocline had the same rate
of rotation, slightly faster than the core rotation rate, in good
agreement with observations. This rate was found to be the same through
several cycles. Again, the model with near-surface $\alpha$ was different, with 
a rotation rate slower than the core rotation. 

The spatial distribution of the non-axisymmetric toroidal fields was
found to be concentrated  around $30^{\circ}$ latitude,
where the radial gradient of the differential 
rotation vanishes. [The latitude, at which the axisymmetric mode is 
concentrated is higher. However, in axisymmetric dynamos, it is possible to vary the meridional circulation and a latitudinal distribution of $\alpha$ to lower the latitudinal distribution of toroidal fields \cite{RueBran95,DikGil01}. ]
When $\alpha$ was finite near the tachocline, the non-axisymmetric
toroidal field had a strong component there. A near surface
concentration was also found for all three models. 

A preliminary study of meridional circulation indicated that the
overall field distribution is not strongly affected, particularly for the toroidal
non-axisymmetric component of the field.  A bigger effect was found in
the cycle period, which becomes longer at higher flow intensities, and
the equatorial symmetry of the solutions, which becomes quadrupolar
(symmetric) when flow intensity is very high. 

In a forthcoming study we will investigate the role of
meridional circulation in the non-axisymmetric dynamo in more details . 
%~~~~~~~~~~~~~~~~~~~~~~~~~~~~~~~~~~~~~~~~~~~~~~~~~~~~~~~~~~~~~~~~~~~~~~
%\begin{acknowledgements}
\section*{Acknowledgements}
%~~~~~~~~~~~~~~~~~~~~~~~~~~~~~~~~~~~~~~~~~~~~~~~~~~~~~~~~~~~~~~~~~~~~~~
We are grateful to Axel Brandenburg for his contribution 
in the developing of the numerical
code. We thank Joan Feynman for her careful examination of the 
manuscript and her advices on the relation between our work and the
solar data,  and to Marco Velli for helpful discussions.  We thank the reviewer for useful critical comments that resulted in improvement of our paper.
This research was conducted in part at the Jet Propulsion Laboratory,
California Institute of Technology, under contract with the National
Aeronautic and Space Administration.
This work was performed while A. Bigazzi held a National Research Council
Research Associateship Award at the Jet Propulsion Laboratory.
%\end{acknowledgements}

%%%%           APPENDICES
%~~~~~~~~~~~~~~~~~~~~~~~~~~~~~~~~~~~~~~~~~~~~~~~~~~~~~~~~~~~~~~~~~~~~~~
\appendix
%~~~~~~~~~~~~~~~~~~~~~~~~~~~~~~~~~~~~~~~~~~~~~~~~~~~~~~~~~~~~~~~~~~~~~~
\section{Description of sources} 
\label{a_prof}
%~~~~~~~~~~~~~~~~~~~~~~~~~~~~~~~~~~~~~~~~~~~~~~~~~~~~~~~~~~~~~~~~~~~~~~
\subsection{Rotation}
%~~~~~~~~~~~~~~~~~ Rotation ~~~~~~~~~~~~~~~~~~~~~~~~~~~~
The differential rotation is approximated by 
\EQ
\label{eq_omega}
\Omega(r,\theta)= {\Omega_0 \over \Omega_{30} }
\bigg(
\Omega_{30} \big(1-f(r)\big) + \Omega_{\mbox{s}} f(r)
\bigg)
,\qquad
f(r) = \frac{1}{2}
\left(
 1 +\tanh\left({r-r_{\Omega}\over\delta_{\Omega}} 
\right)
\right).
\EN
The parameters $r_{\Omega}=0.69$ and $\delta_{\Omega}=0.05$ define the
location and the width of the tachocline.
Here,  
$\Omega_{\mbox{s}}$ is the surface rotation  profile, 
\begin{equation}
\Omega_{\mbox{s}}(\theta)=
1+a_1\cos^2\theta+a_2\cos^4\theta,
\end{equation}
with $a_1=-0.126$, $a_2=-0.159$, 
, and 
$\theta$ is the co-latitude  
\cite{moss99,charbonneau99}.
The equatorial  rotation rate is  thus
given by
\EQ
\Omega_{eq} ={\Omega_0\over\Omega_{30}}
\EN 
where $ \Omega_{30} = \Omega_{\mbox{s}}(30^{\circ})$.
In all runs we use and $\Omega_0 = 4.9\times 10^4 $, in non-dimensional
units. 

\subsection{Diffusivity}
%~~~~~~~~~~~~~~~~~ Diffusivity  ~~~~~~~~~~~~~~~~~~~~~~~~~~~~
The turbulent  diffusivity profile is given by
\EQ
\eta(r)= \eta_c \big(1-f(r)\big) +\eta_0 f(r)
,\qquad
f(r) = {1\over 2} 
\left(
 1 +\tanh\left(
{r-r_{\eta}\over\delta_{\eta}} 
\right)
\right). 
%  prof=0.5*(1+tanh((xx-xeta)/thketa*1.31))
\EN
Here, $r_{\eta}\equiv r_\Omega =0.69$, $\delta_{\eta}=0.04$;   
$\eta_c$ and $\eta_0$ are the diffusivity values in the core and 
in the convection zone respectively. We use  $\eta_0/\eta_c=50$ 
and  $\eta_0/\eta_c=200$.

\subsection{$\alpha$-effect}
%~~~~~~~~~~~~~~~~~ Diffusivity  ~~~~~~~~~~~~~~~~~~~~~~~~~~~~
%~~~~~~~~~~~~~~~~~ alpha ~~~~~~~~~~~~~~~~~~~~~~~~~~~~
The radial dependences of the $\alpha$-effect for the three cases considered, 
are: 
\EQ
\alpha(r)=\alpha_0 \; {1 \over 2} 
\left(
1 +\tanh{r-r_{\alpha}\over\delta_{\alpha}} 
\right)
\EN
with  $\alpha_0=0.9$, $r_{\alpha}=0.7$, $\delta_{\alpha}=0.03$, in model M1,
and 
 $\alpha_0=24.2$, $r_{\alpha}=0.9$, $\delta_{\alpha}=0.03$ in model M2; 
%
    %alpr=.5*(1.+tanh((xx-xlef)/dalp)) *  .5*(1.-tanh((xx-xrig)/dalp))
\EQ
\alpha(r)=\alpha_0 \; 
{1 \over 2} 
\left(
1 +\tanh{r-r_{\alpha}\over\delta_{\alpha}} 
\right)
\cdot
{1 \over 2} 
\left(
1 -\tanh{r-r_{\alpha}+0.07\over\delta_{\alpha}} 
\right)
\EN
with  $\alpha_0=1.9$, $r_{\alpha}=0.75$, $\delta_{\alpha}=0.04$, for model
M3. 
The values of $\alpha_0$ have been chosen so that the 
dynamo is supercritical but  close to marginal. 
%~~~~~~~~~~~~~~~~~~~~~~~~~~~~~~~~~~~~~~~~~~~~~~~~~~~~~~~~~~~~~~~~~~~~~~
\subsection{Meridional circulation}

\begin{figure}
%\figurenum{<text>}
\epsscale{1.5}
%\plotone{f3.ps}
\plottwo{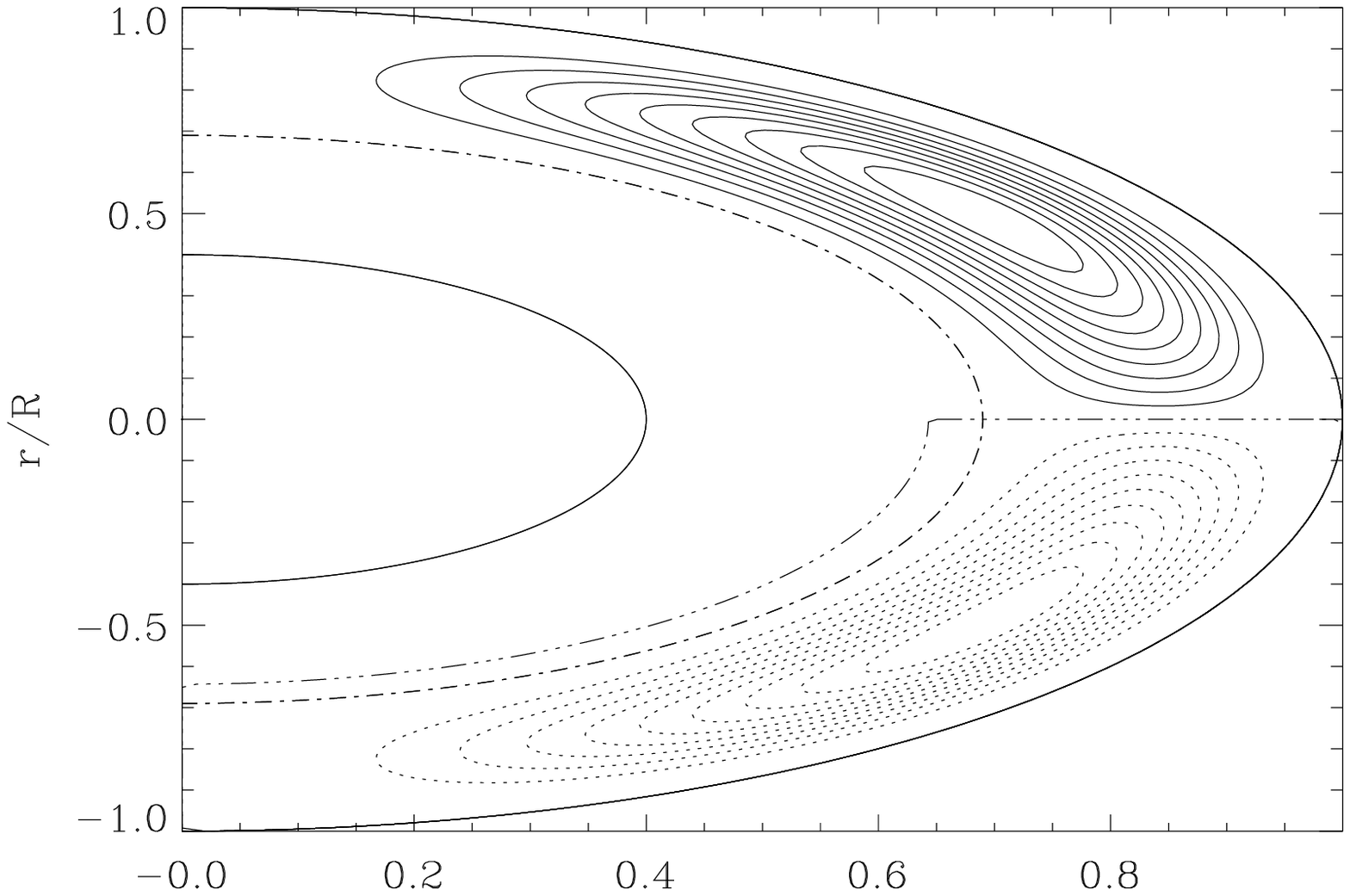}{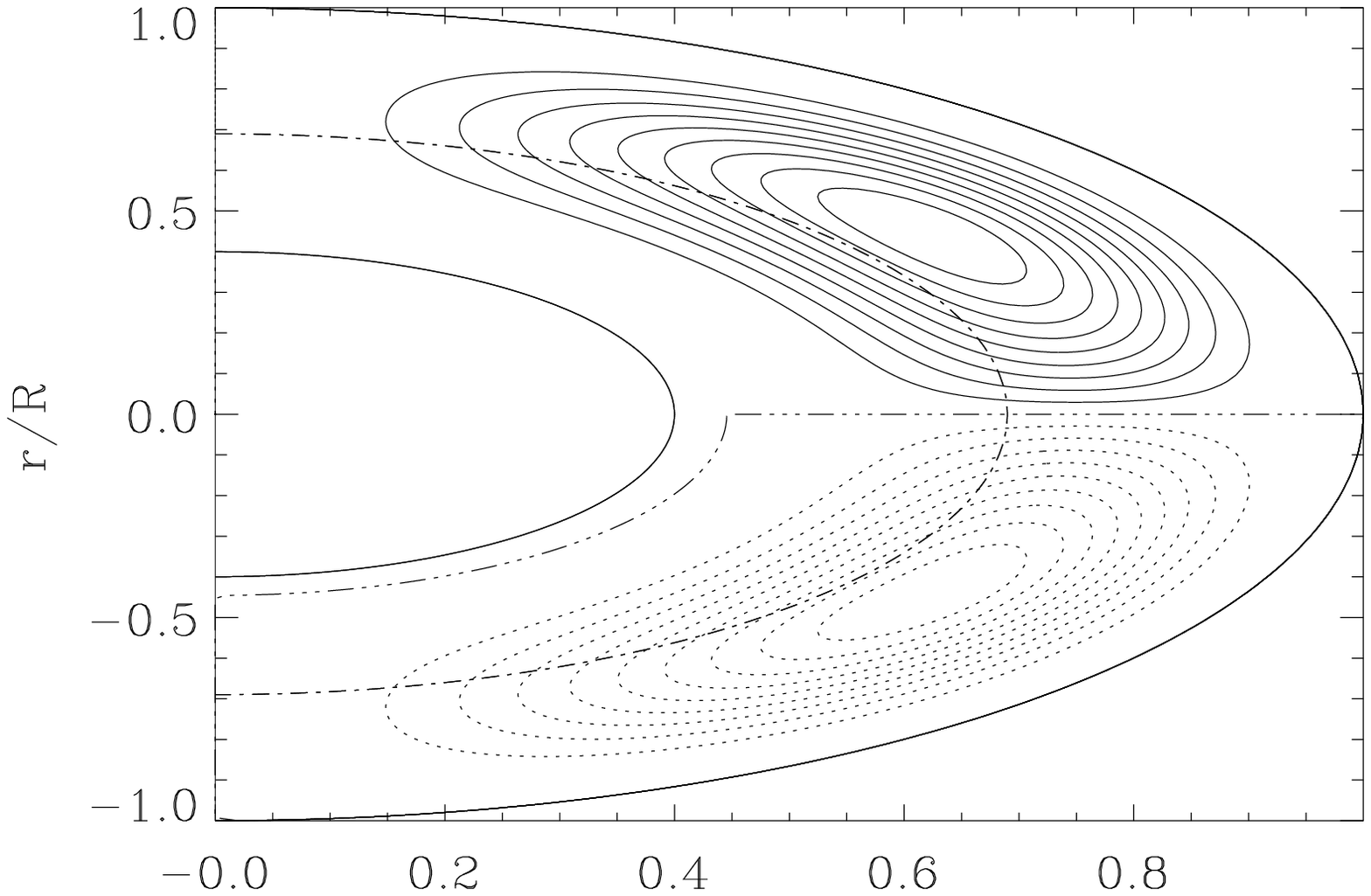}
\caption{
Meridional circulation. Flow lines for the two cases considered. 
Solid lines  indicate anti-clockwise flow and dotted lines clockwise. 
The dashed line at $0.69 R_{\odot}$ indicates the centre of the tachocline. 
\label{fig_merid}
}
\end{figure}

We consider a simple 2-cell flow 
where the surface flow amplitude  is of the order of that observed. 
%We consider M1. 
Density stratification in the convection zone is very high, therefore 
mass continuity has to be taken into account. 
We specify a simple fit to reproduce the density from a  solar model 
\cite{Christ96},
between $0.60  R_{\odot}$ and  $0.95  R_{\odot}$.
\EQ
\rho={\alpha e^{-\beta r}}
\EN
where $\alpha= 247$ and $\beta= 10.1$.
We impose mass continuity
\EQ
\nab \rho \uu =0
\EN
by introducing  a poloidal potential $\Psi$ for the mass flow:
\EQ
\rho \uu =  \nab \times \nab \times \rr \Psi(r,\theta).
\EN
The analogous Stokes stream function is
$- r \sin\theta  \; \partial_{\theta} \Psi$.
Streamlines for this flow are contours where such a function is constant.
Fig.~\ref{fig_merid} shows the two cases of the meridional 
circulation where the flow
does or does not penetrate inside the shear region (tachocline).
Those  profiles result from a simple power-law expression
\cite{RobStix72}
for the radial part of the 
potential:
\EQA
\Psi(r)&=& \frac{2}{(1-r_b)^6} (r-r_b)^3 (1-r)^3  \qquad r>r_b
\nonumber \\
\Psi(r)&=&0 \qquad r < r_b,
\ENA
where $r_b=0.45$ for the deep flow case and $r_b=0.65$ for the shallower
case.
We specify the latitudinal dependence to be proportional to 
\EQ
 P_{2,0}(\cos \theta) \sim  3\cos^2 \theta -1 
\EN
where $P_{l,m}$ is the associated Legendre polynomial of order $l=2$ and degree
$m=0$. This splits the domain into two counter-rotating cells, each poleward
of the equator. 

%~~~~~~~~~~~~~~~~~~~~~~~~~~~~~~~~~~~~~~~~~~~~~~~~~~~~~~~~~~~~~~~~~~~~~~
%%%%            BIBLIOGRAPHY
%~~~~~~~~~~~~~~~~~~~~~~~~~~~~~~~~~~~~~~~~~~~~~~~~~~~~~~~~~~~~~~~~~~~~~~

%
%

{}

\clearpage

                         %%% FIGURES %%%
%--------------------------------------------------
\begin{figure}
%\figurenum{<text>}
\epsscale{1.}
%\plotone{./fig/omega_prof.eps}
\plotone{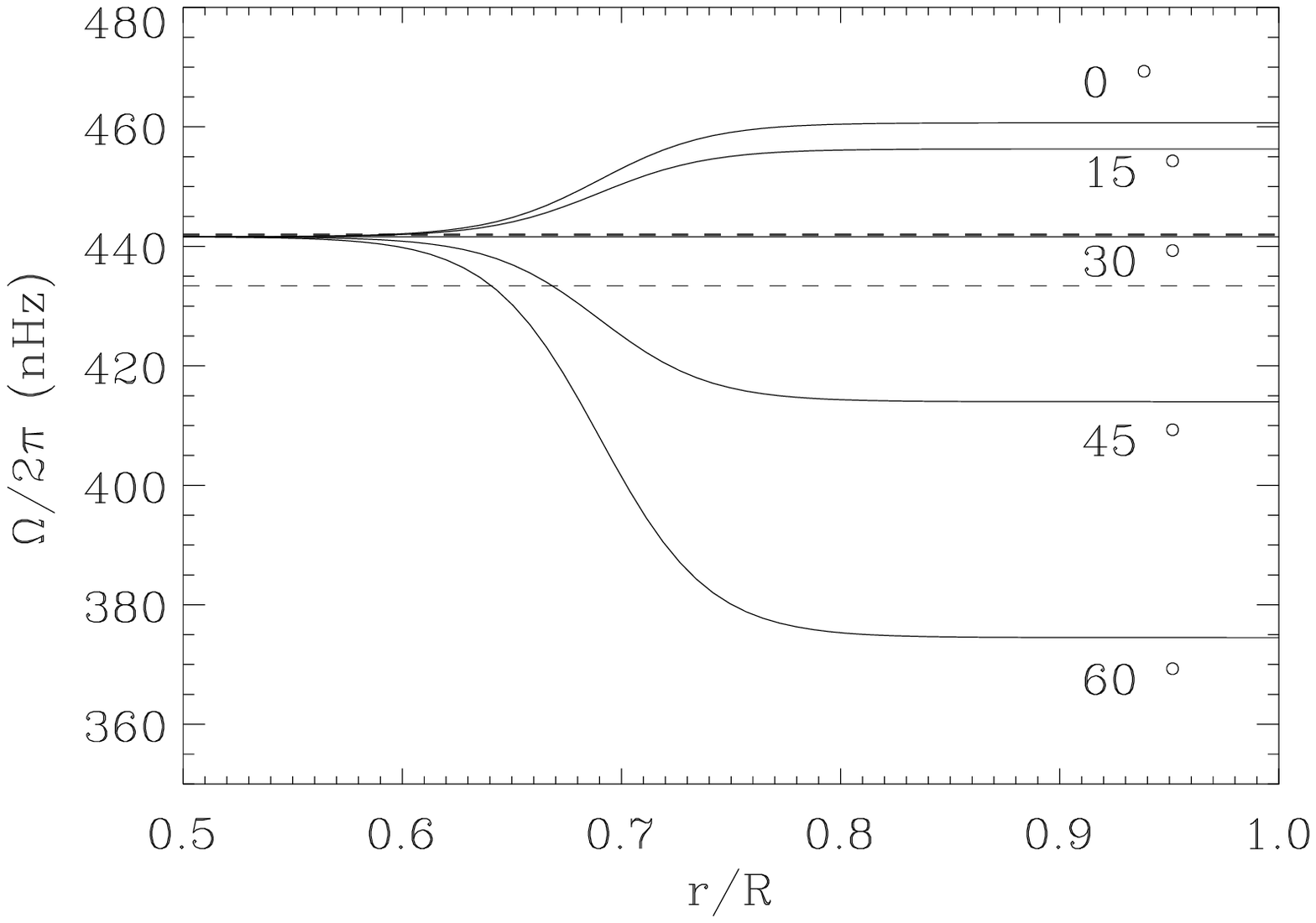}
%\plottwo{<epsfile>}{<epsfile>}
\caption{
Distribution of the internal solar rotation used in our dynamo
modeling.
Dashed curves show the rotation rates of the non-axisymmetric mode, see 
Table~(\ref{tab_result}).
The upper curve at $442 {\rm nHz}$, just above the core rotation rate, 
  is for models M1 and M3. 
The lower curve, at $433  {\rm nHz}$,  is for model  M2. 
\label{fig_rotOurs}
}
\end{figure}
%-------------        -----------------------------
\begin{figure}
%\figurenum{<text>}
\epsscale{1.}
\plotone{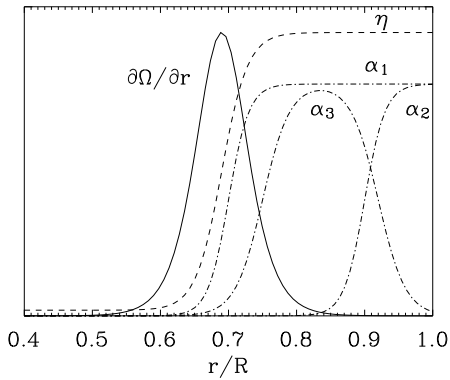}
%\plottwo{<epsfile>}{<epsfile>}
\caption{
Radial distributions of the sources  of magnetic field generation:
$\alpha$, for three different models,  the radial gradient
of the rotation rate at the equator 
%$\partial\Omega/\partial r$ 
and turbulent diffusivity $\eta$.
Values are not to scale;  for actual values see Appendix. 
\label{fig_alp}
}
\end{figure}
%-------------        -----------------------------
%--------------------------------------------------
%\clearpage
%--------------------------------------------------
%-------------  PHASE0      -----------------------------
%-------------        -----------------------------
%-------------  PHASE2      -----------------------------
%-------------  PHASE3      -----------------------------
%--------------------------------------------------
%--------------------------------------------------
\onecolumn
%--------------------------------------------------
%---------------Field Distr -----------------------------------
%\clearpage
%\clearpage
%--------------------------------------------------
%-------------------------------------------------- BUTFLYS
%\clearpage
%--------------------------------------------------
	%--------------------------------------------------
	%--------------------------------------------------
	%--------------------------------------------------
	%--------------------------------------------------
	%--------------------------------------------------
	%--------------------------------------------------
	%--------------------------------------------------

\clearpage
                         %%% TABLES %%%
%--------------------- TABLE Results  -------------------------------
\begin{table}
%\begin{tabular}{lrrrrrrrrr}
\begin{tabular}{lrrr}
%-----------------------------------------------------------------------------
\hline
Model  & M1 & M2  & M3    \\
\hline
 $\gamma$		& 0.35	& 4.15 	  & 1.4   \\
 $T$			& 0.022& 0.0026   	& 0.018 	  \\
 $T/T_{rot}$		&  179	&  20.2	  	& 137 	  \\
 $\Omega^1$ (nHz)        & 442 & 433  	& 442 	  \\
\hline
\end{tabular}
\caption{Time evolution.
\label{tab_result} } 
%\caption{ Growth rates and cycle periods for the three
%models M1, M2 and M3.  \label{tab_result} } 
\end{table}

%\clearpage

%------------------ Table phase ----------------------------------
\begin{table}
\begin{tabular}{lrrr}
\hline
%-----------------------------------------------------------------------------
  Model & M1 & M2 & M3  \\
\hline
$\Delta \varphi$ ($B_{\varphi}
\rightarrow B_r$) & $1/6 \pi$ & $ -3/4 \pi$ & $ 1/2 \pi$ \\ 
$\Delta \varphi$ ($S_0 \rightarrow S_1$)   & 
        $1.1 \pi$  & $ 1.4 \pi$  & $1.1 \pi$  \\
\hline
\end{tabular}
\caption{Phase relations.
%\caption{ Phase lag between field components
%and between 
%modes of the poloidal potential.
\label{tab_phase}
}
\end{table}
%--------------------- TABLE   -------------------------------

%\clearpage

\begin{table}
\begin{tabular}{lrrr}
%-----------------------------------------------------------------------------
\hline
Model  & M1 & M2  & M3    \\
\hline
 $r_M^0$ 		&0.70  & 0.95  	& 0.72 ; 0.90  	  \\
 $\theta_M^0$(lat) 	& 57 & 50    & 60 ; 52 	  \\
$r_M^1$ 		&0.67 ; 0.93 & 0.90 ; 0.97 & 0.78 ; 0.93  	\\
 $\theta_M^1$(lat)      & 30 ; 33           & 42 ; 41    & 32 ; 32  \\
% $\delta\Omega_1$       & 26& -910  	& 48 	  \\
% $\Omega^1$ (nHz)        & 442 & 433  	& 442 	  \\
\hline
\end{tabular}
\caption{
Location of magnetic field maxima.
%Locations of the maxima of the toroidal field distributions of
%Figs.~\ref{fig_toro_m0m1-1}-\ref{fig_toro_m0m1-3}; rotation rate
%of the $m=1$ mode (see Fig.~\ref{fig_magn_m1+phase}).
\label{tab_distr}
}
\end{table}

%\clearpage

%--------------------- TABLE   -------------------------------
\begin{table}
\begin{tabular}{lrr}
%-----------------------------------------------------------------------------
\hline
Model  & M1S & M1D    \\
\hline
 $r_M^0$ 		&0.80  & 0.65  	 	  \\
 $\theta_M^0$(lat) 	& 55 & 59  	  \\
$r_M^1$ 		&0.67 ; 0.90 & 0.65 ; 0.87  	\\
 $\theta_M^1$(lat)      & 31 ; 34           & 30 ; 35  \\
\hline
\end{tabular}
\caption{
Location of magnetic field maxima.
%Same as in Table~\ref{tab_distr}, for model M1 with shallow (M1S) and
%deep (M1D) meridional circulation (see Figs.~\ref{fig_toro_m0m1-1ms}
%and \ref{fig_toro_m0m1-1md}).  
\label{tab_merid} }
\end{table}
%%%%%%%%%%%%%%%%%%%%%%%%%%%%%%%%%%%%%%%%%%%%%%%%%%%%%%%%%%%%%%%%%%%%%%%%%

%%%%%%%%%%%%%%%				       %%%%%%%%%%%%%%%%%%%%%%%%%%
\end{document}